\documentclass[12pt,a4paper]{article}
\usepackage{amsmath,amssymb,graphicx,epsfig,cite,color}

\usepackage[left=2.7cm,right=2.7cm,top=2.5cm,bottom=2.5cm]{geometry}
\usepackage{subfigure}
\usepackage{multirow}
\usepackage[table]{xcolor}
\usepackage{booktabs}

\usepackage{xcolor,colortbl}

\usepackage{graphicx}
\usepackage{lscape}
\usepackage{appendix}
\usepackage{multirow}
\definecolor{DarkPastelGreen}{rgb}{0.01,0.69,0.28}
\definecolor{amethyst}{rgb}{0.6, 0.4, 0.8}
\definecolor{darkcyan}{rgb}{0.0, 0.55, 0.55}
\usepackage[colorlinks=true,linkcolor=amethyst,citecolor=darkcyan,urlcolor  = DarkPastelGreen]{hyperref}%
\allowdisplaybreaks[1]

\begin{document}

\def\ds{\displaystyle}
\def\beq{\begin{equation}}
\def\eeq{\end{equation}}
\def\bea{\begin{eqnarray}}
\def\eea{\end{eqnarray}}
\def\beeq{\begin{eqnarray}}
\def\eeeq{\end{eqnarray}}

\def\rar{\rightarrow} 
\def\nnb{\nonumber}

\def\ds{\displaystyle}
\def\beq{\begin{equation}}
\def\eeq{\end{equation}}
\def\bea{\begin{eqnarray}}
\def\eea{\end{eqnarray}}
\def\beeq{\begin{eqnarray}}
\def\eeeq{\end{eqnarray}}
\def\ve{\vert}
\def\vel{\left|}
\def\ver{\right|}
\def\nnb{\nonumber}
\def\ga{\left(}
\def\dr{\right)}
\def\aga{\left\{}
\def\adr{\right\}}
\def\lla{\left<}
\def\rra{\right>}
\def\rar{\rightarrow}
\def\lrar{\leftrightarrow}
\def\nnb{\nonumber}
\def\la{\langle}
\def\ra{\rangle}
\def\ba{\begin{array}}
\def\ea{\end{array}}
\def\tr{\mbox{Tr}}
\def\ssp{{\Sigma^{*+}}}
\def\sso{{\Sigma^{*0}}}
\def\ssm{{\Sigma^{*-}}}
\def\xis0{{\Xi^*}}
\def\xism{{\Xi^{*-}}}
\def\qs{\la \bar s s \ra}
\def\qu{\la \bar u u \ra}
\def\qd{\la \bar d d \ra}
\def\qq{\la \bar q q \ra}
\def\gGgG{\la g^2 G^2 \ra}
\def\q{\gamma_5 \not\!q}
\def\x{\gamma_5 \not\!x}
\def\g5{\gamma_5}
\def\sb{S_Q^{cf}}
\def\sd{S_d^{be}}
\def\su{S_u^{ad}}
\def\sbp{{S}_Q^{'cf}}
\def\sdp{{S}_d^{'be}}
\def\sup{{S}_u^{'ad}}
\def\ssp{{S}_s^{'??}}

\def\sig{\sigma_{\mu \nu} \gamma_5 p^\mu q^\nu}
\def\fo{f_0(\frac{s_0}{M^2})}
\def\ffi{f_1(\frac{s_0}{M^2})}
\def\fii{f_2(\frac{s_0}{M^2})}
\def\O{{\cal O}}
\def\sl{{\Sigma^0 \Lambda}}
\def\es{\!\!\! &=& \!\!\!}
\def\ap{\!\!\! &\approx& \!\!\!}
\def\ar{&+& \!\!\!}
\def\ek{&-& \!\!\!}
\def\kek{\!\!\!&-& \!\!\!}
\def\cp{&\times& \!\!\!}
\def\se{\!\!\! &\simeq& \!\!\!}
\def\eqv{&\equiv& \!\!\!}
\def\kpm{&\pm& \!\!\!}
\def\kmp{&\mp& \!\!\!}
\def\mcdot{\!\cdot\!}
\def\erar{&\rightarrow&}

% .........................................................

% .........................................................

\renewcommand{\textfraction}{0.2}    %float (figures) parameters
\renewcommand{\topfraction}{0.8}   

\renewcommand{\bottomfraction}{0.4}   
\renewcommand{\floatpagefraction}{0.8}
\newcommand\mysection{\setcounter{equation}{0}\section}

\def\baeq{\begin{appeq}}     \def\eaeq{\end{appeq}}  
\def\baeeq{\begin{appeeq}}   \def\eaeeq{\end{appeeq}}
\newenvironment{appeq}{\beq}{\eeq}   
\newenvironment{appeeq}{\beeq}{\eeeq}
\def\bAPP#1#2{
 \markright{APPENDIX #1}
 \addcontentsline{toc}{section}{Appendix #1: #2}
 \medskip
 \medskip
 \begin{center}      {\bf\LARGE Appendix #1 :}{\quad\Large\bf #2}
% \begin{center}      {\bf\LARGE Appendix  :}{\quad\Large\bf #2}
\end{center}
 \renewcommand{\thesection}{#1.\arabic{section}}
\setcounter{equation}{0}
        \renewcommand{\thehran}{#1.\arabic{hran}}
\renewenvironment{appeq}
  {  \renewcommand{\theequation}{#1.\arabic{equation}}
     \beq
  }{\eeq}
\renewenvironment{appeeq}
  {  \renewcommand{\theequation}{#1.\arabic{equation}}
     \beeq
  }{\eeeq}
\nopagebreak \noindent}

\def\eAPP{\renewcommand{\thehran}{\thesection.\arabic{hran}}}

\renewcommand{\theequation}{\arabic{equation}}
\newcounter{hran}
\renewcommand{\thehran}{\thesection.\arabic{hran}}

\def\bmini{\setcounter{hran}{\value{equation}}
\refstepcounter{hran}\setcounter{equation}{0}
\renewcommand{\theequation}{\thehran\alph{equation}}\begin{eqnarray}}
\def\bminiG#1{\setcounter{hran}{\value{equation}}
\refstepcounter{hran}\setcounter{equation}{-1}
\renewcommand{\theequation}{\thehran\alph{equation}}
\refstepcounter{equation}\label{#1}\begin{eqnarray}}

%       the stuff below defines \eqalign and \eqalignno in such a
%       way that they will run on Latex

\newskip\humongous \humongous=0pt plus 1000pt minus 1000pt
\def\caja{\mathsurround=0pt}
%\def\eqalign#1{\,\vcenter{\openup1\jot
%\caja   %\ialign{\strut \hfil$\displaystyle{##}$&$
%\displaystyle{{}##}$\hfil\crcr#1\crcr}
%}\,}
 
% ...........................................................

\title{
         {\Large
                 {\bf
Meson mass and  width: Deep learning approach
                 }
         }
      }

\author{\vspace{1cm}\\
	{\small
	M.  Malekhosseini$^{1}$,
	S.~Rostami$^{1}$,
		A. R. Olamaei$^{2,3}$,
		R. Ostovar$^{4}$,
		K. Azizi$^{1,5,6}$
		\thanks{Corresponding author}
		}
\\
	{\small $^1$ Department of Physics, University of Tehran, North Karegar Avenue, Tehran 14395-547, Iran}\\
	{\small $^2$ Department of Physics, Jahrom University, Jahrom, P.  O.  Box 74137-66171, Iran}\\
	{\small$^3$  School of Physics, Institute for Research in Fundamental Sciences (IPM),}\\
                         {\small  P. O. Box 19395-5531, Tehran, Iran}\\
                     {\small$^4$    School of Cognitive Sciences, Institute for Research in Fundamental Sciences (IPM),}\\
                       {\small P. O. Box 19568-36489, Tehran, Iran,}\\
	{\small $^5$ Department of Physics, Do\u gu\c s University, Dudullu-\"Umraniye, 34775 Istanbul, T\"urkiye}\\
	{\small $^6$ Department of Physics and Technical Sciences, Western Caspian University,  Baku, AZ 1001, Azerbaijan}
	} 

\date{}

\begin{titlepage}
\maketitle
\thispagestyle{empty}

\begin{abstract}

It is fascinating to predict the  mass and  width of the ordinary and exotic mesons solely based on their quark content and quantum numbers. Such prediction goes beyond conventional methodologies traditionally employed in hadron physics for calculating or estimating these quantities.  The relation between the quantum numbers and the properties of the mesons, such as the mass and width, is complicated in the world of particle physics. 
However, the deep neural network (DNN) as a subfield of machine learning techniques
 provides a solution to this problem. By analyzing large datasets, deep learning algorithms
 can automatically identify complex patterns among the particles' quantum numbers, and their  mass and width, 
 that would otherwise require complex calculations.  
In this study, we present two approaches using the DNNs
 to estimate the mass of some ordinary and exotic mesons. 
Also for the first time, the DNNs are trained to predict the  width of ordinary and exotic mesons, 
whose widths have not been experimentally known.
Our predictions obtained through the DNNs, will be useful for future experimental searches.
\end{abstract}
 
%\vspace{1cm}
%~~~PACS number(s): 11.55.Hx, 13.75.Gx, 13.75.Jz
\end{titlepage}
\section{Introduction}
Nature suggests that hadrons are composite subatomic particles containing two or more quarks binding together via strong interaction.
The realm  of hadrons is divided into baryons and mesons. Baryons are composed  of three valence-quarks and mesons are created as  valence quark-antiquark bound states \cite{Gell-Mann:1964ewy,Gell-Mann:1962yej,Chodos:1974je}.
As we know, the true structure of a hadron may be more complex than a simple composition of two or three quarks \cite{Griffiths:1983ah,Chen:2022asf}. 
Obtaining the quantum numbers and physical structures of these
 hadronic subatomic particles provides us with important pieces of information. 
Also, the mass and width are two determinative quantities for theorists 
and experimentalists' researches. The origin of the hadron mass has been a challenging topic 
 in particle physics for a long time. 
 The mass of a hadron is influenced by not only the quark 
 content but also the dynamics within the particle. Indeed, 
 the creation and annihilation of virtual quark-antiquark  pairs and virtual gluons
  and their interactions with the QCD vacuum (condensations)
 contribute to the hadronic compositions \cite{Schumacher:2018evl,Schumacher:2015wla}.
 According to the quark model, the hadrons going beyond $ q\bar{q} $ and  $ qqq $ compositions are 
 called exotic hadrons. For instance, the four-quark mesons or tetraquarks 
 are composed of  $q\bar{q}q\bar{q} $. 
 These exotic states have been explored by
  theorists \cite{Jaffe:1976ig,Jaffe:1976ih,Dong:2019ofp,Voloshin:1976ap,Esposito:2016noz,Shifman:1978bx,Agaev:2016dev,Agaev:2016srl,Agaev:2020zad,Azizi:2019xla} and observed by the LHCb and other Collaborations \cite{Cowan:2016kjn,LHCb:2015yax,LHCb:2016lve,Guo:2015umn,LHCb:2022sfr,LHCb:2016axx,CDF:2003cab,Belle:2003nnu,CMS:2023owd,LHCb:2021vvq,LHCb:2021auc}. 
In recent years, various kinds of  exotic particles have been studied through 
theoretical methods such as  QCD sum rules (QSR) \cite{Choe:1996uc,Agaev:2023tzi,Agaev:2023wua,Agaev:2023ruu}, potential models and lattice QCD \cite{HadronSpectrum:2012gic,Dudek:2009qf}. ATLAS experiment has presented results on exotic resonances from the proton-proton collision \cite{Bouhova-Thacker:2022vnt}. 

To date, the Particle Data Group (PDG) \cite{ParticleDataGroup:2022pth} has documented 
hundreds  mesonic and  baryonic states.
 After all, it is still a hot topic to probe
structures of the hadrons and exotic states, since the internal hadronic configurations
of quark and gluons can be obtained in high-energy reactions \cite{VanHove:1974wa}, thanks to the
progresses made at different hadron colliders.  
 Moreover, it is 
important to highlight that the new exotic particles reveal our limited knowledge 
 of the hadronic systems due to the gap between the theory and experiment \cite{Briceno:2015rlt}.

Artificial intelligence and the machine learning  (AI/ML) techniques have the 
potential to bridge this gap and aid  the theory and experiment for improved  
performance and accuracy \cite{Calafiura:2022ges,Schwartz:2021ftp}. 
The traditional AI and ML methods have been used in high-energy physics (HEP) since the 1990s \cite{Calafiura:2022ges,Schwartz:2021ftp,Guest:2018yhq,Albertsson:2018maf}. 
Afterward, the ML approaches were mostly applied in particle and event identification as well as reconstruction in the 2010s, and consequently played a role in discovery of the Higgs boson at the Large Hadron Collider (LHC) in 2012  \cite{Calafiura:2022ges,Schwartz:2021ftp,Guest:2018yhq,Albertsson:2018maf,Duarte:2022job,CMS:2012qbp,CMS:2018nsn,ATLAS:2012yve}. 

The recent breakthroughs in the modern ML 
have been transfiguring particle physics studies in both 
phenomenological and experimental areas. In other words, 
the modern ML facilitates collaboration between LHC experimentalists, 
phenomenologists, and the data science community  \cite{Calafiura:2022ges,Schwartz:2021ftp,Guest:2018yhq,Albertsson:2018maf,Duarte:2022job,CMS:2012qbp,CMS:2018nsn,ATLAS:2012yve,Larkoski:2017jix,Radovic:2018dip,Carleo:2019ptp,Bourilkov:2019yoi,Feickert:2021ajf,Karagiorgi:2021ngt,Shanahan:2022ifi,Plehn:2022ftl,Butter:2022rso}. Some established ML algorithms, 
such as boosted decision trees (BDTs) and neural networks (NNs), have a 
long and successful history in many experimental analyses. These powerful 
methods are now considered standard tools and are extensively implemented by the ATLAS, CMS, and LHCb Collaborations \cite{Plehn:2022ftl,Butter:2022rso,CMS:2017wtu,ATLAS:2015yey,Pata:2022wam,Stoye:2018qgr,Wachirapusitan:2023ttc,Schramm:2018knx,DeCian:2017ytk,ATLAS:2017gpy,Keicher:2023mer,ATLAS:2023ixc,ATLAS:2015jge,Graczykowski:2022zae}. In the first observation of the Higgs boson, 
the ATLAS Collaboration successfully leveraged  
the NN techniques \cite{ATLAS:2012yve,Schramm:2018knx}. Similarly, the LHCb Collaboration used the ML tools for the first observations of the exotic structures referring to charmonium-pentaquark states \cite{LHCb:2015yax,Schramm:2018knx}.

The NNs are a branch of the ML models that use learning algorithms inspired by the human brain
to recognize complex relations between various parameters based on large numbers of examples
and make intelligent decisions.
The NNs receive raw input data in the first layer (input layer), process it through at least 
one hidden layer, and then present the result in the last layer (output layer). 
The simplest form of a NN consists of an input layer, a hidden layer, and an output layer.
Deep neural networks (DNNs) are more advanced, with several hidden layers. These networks are capable of solving more complex problems 
\cite{Bas,Roberts:2021fes}.
Application of the NNs to the HEP experiments and phenomenology has been very successful 
 \cite{Guest:2018yhq,Alison:2019kud,Chen:2022ddj,Andrews:2019faz,Zhang:2023czx,Ng:2021ibr,ExaTrkX:2020nyf,Bahtiyar:2022une,Gal:2020dyc,Zhang:2024bld,Bogatskiy:2023nnw,Athanassopoulos:2003qe}.
The DNNs are becoming indispensable tools in searching and unraveling dark matter nature 
 \cite{Arganda:2024eub,Chacon:2023nhw,Thais:2022iok}. 
 Graph neural network (GNN) models are a further extension of NNs, able to process data in graph structures. The GNNs show great promise in particle reconstruction problems
 \cite{ExaTrkX:2020nyf}. 
 One of the most powerful types of deep learning models is Generative Adversarial Network (GAN). A GAN is a type of artificial intelligence framework that consists of two NNs, a generator and a discriminator which compete with each other via deep learning algorithms. They are trained simultaneously through adversarial training. The generator network generates new data samples,  while the discriminator network tries to distinguish between real data samples and fake ones generated by the generator. Event generation,  anomaly detection, data augmentation, and background subtraction are just a few examples of how the GANs can be applied in the field of particle physics to empower data analysis, simulation, and detection capabilities \cite{Butter:2022rso,Paganini:2017dwg,Krause:2021ilc,Butter:2021csz,Heimel:2022wyj,Hashemi:2023ruu}.
 
 The DNNs are standard tools for solving classification and regression problems 
 \cite{Alison:2019kud,Chen:2022ddj,Andrews:2019faz,Zhang:2023czx,Ng:2021ibr,ExaTrkX:2020nyf,Bahtiyar:2022une,Gal:2020dyc,Zhang:2024bld,Bogatskiy:2023nnw,Athanassopoulos:2003qe}. 
 In the classification tasks, the output of the algorithm is restricted to specific values, or classes, such as ``signal" or ``background" 
 \cite{Feickert:2021ajf,Butter:2022rso}. 
 An innovative DNN method has been represented for jet identification in the CMS experiment at 
the LHC \cite{Alison:2019kud}. Ref. \cite{Chen:2022ddj} trained and 
developed the DNNs to judge whether the exotic states of the $ X(3872) $, $ X(4260) $ and $ Z_c(3900) $ can be considered as hadronic molecules of special channels. 
In addition, the nature of hidden charm pentaquarks with a NN approach was studied \cite{Zhang:2023czx}. 
As previously explained, the NNs have the potential to perform the regression 
analysis and predict continuous numerical values. The regression algorithms can
 be implemented for reconstruction techniques in the HEP, e.g. precise calculations 
 of continuous quantities like, hit positions and track momenta or the jet energies \cite{Feickert:2021ajf,Butter:2022rso,CMS:2017wtu,Stoye:2018qgr,Baldi:2018qhe,Karagiorgi:2022qnh}.
The ATLAS collaboration has employed a regression DNN on a sample of simulated $ t\bar{t} $ 
  events to reconstruct the top pair system invariant mass \cite{Guerrieri:2022bxg}. 
  Also, the NNs can be utilized to compute the mass of the exotic hadrons, doubly charmed and bottom baryons.
   For this purpose, the original data was extended by 
  using artificial data augmentation methods \cite{Bahtiyar:2022une}. Interestingly, 
  Ref. \cite{Gal:2020dyc} uses basic information of the mesons spectrum to predict 
  the masses of the baryons, pentaquarks and other exotic hadrons. Inspired by these 
  applications and based on the search strategies performed in \cite{Gal:2020dyc}, 
  it is a fortunate time to extend this scheme and design the DNNs to precisely 
  compute the mass and width of some ordinary and exotic mesons.
The quark content of some famous mesons are still unclear and
their identification has remained a complicated experimental task.
We estimate the mass of these challenging mesons through the DNNs, 
considering the ordinary and exotic structures for their quark contents separately. 
Furthermore, we examine the power of our designed DNNs in prediction of 
the mass of some light and exotic mesons.   
Moreover, the DNNs are implemented to predict 
the decay width of some ordinary and exotic mesons whose decay widths have not been 
permanently confirmed by the PDG \cite{ParticleDataGroup:2022pth}.   
  
In our study, the DNNs are trained based on two effective approaches in organizing the dataset, 
according to the quark contents and global quantum numbers of the meson 
such as the angular momentum, parity, isospin and charge conjugation. 
The obtained results
regarding the mass spectra of the normal and exotic mesons are found to be in good
agreement with the experimental results. 
For the first time, we are able to predict the decay width 
of the mesonic states with a good approximation based on their quark 
contents and the other quantum numbers like the angular momentum, parity, etc. and
also mass a new input, through the deep learning algorithms.
In this analysis, the encoded data of the mesons are given to the DNNs.
The DNN recognizes the mesons in terms of their quark contents and  the other global quantum numbers.
The data structure is examined in each hidden layer. 
Each layers' output is a new description of the input data for the next layer.
Improving this process, layer by layer, until the final predictions of the mass and width are presented.
At last, the DNN converts the encoded data into the practical output.

Learning by example and transferring data at each layer are known as the most powerful aspects of neural network.
Despite the broad deployment of the NNs in strongly correlated systems and pattern recognition tasks, 
it has not been clarified how the individual neurons work together to reach the final output. 
The DNNs are typically considered as one of the most famous black box algorithms in extracting knowledge from data.
We have designed the DNNs which are able to predict the mesons’
mass spectra and widths, whereas, most mesons listed in the PDG are unstable and
determination of their mass and total decay width requires various complex methods in
the theory and experiment. 
This is the point that makes our research more attractive.

The structure of this paper is in the following form. In section \ref{NN}, the DNN is briefly introduced. 
The details of the data preprocessing is described in section \ref{NA}. 
In section \ref{r}, we present and discuss the results of our analyses. 
Lastly, the section \ref{SC} is reserved for our concluding notes.

\section{Neural Networks}\label{NN}
The ML models aim to automatically learn  to recognize complicated patterns and make intelligent decisions based on data. The NNs, also known as artificial NNs (ANNs) are  famous framework to perform ML goals.
 In fact, the NN uses a series of learning algorithms to analyse 
 and function in a manner inspired by the human brain to find patterns among vast amounts of data.

A typical NN is composed of highly interconnected processing elements called neurons or nodes, 
organized in several layers. Indeed, the most fundamental component of the NN is the neuron, 
which takes incoming information from the neurons of the previous layer or directly from the input features. 
After processing the information through some mathematical functions, 
the neuron transmits the output value to the neurons of the
next layer, i.e.  the outputs of the neurons in one layer will be the inputs for the
next layer. 
There exist three types of layers in the NN. 
An input layer, at least one hidden layer and one output layer. 
The input layer is not responsible for any computation or transformation.  
It just passes the data features to the next layers. 
The process of passing input data through the NN in a forward direction,
from the input layer through any hidden layers and finally to the output layer is called feedforward. 
A feedforward NN is one of the most common and successful learning algorithms where 
information moves only in one direction.
Consider a feedforward NN with $ L $ layers, so we have one input layer, one output layer, and $ L-2 $ hidden layer. 
A neuron transforms the input values $\mathbf{x}$ into the output $\mathbf{y}$,

\begin{equation}
\textbf{x} \rightarrow \textbf{x}^{1}\rightarrow\textbf{x}^{2} . . .\rightarrow \textbf{x}^{L}\equiv  \textbf{y},
\end{equation}
where  $ \textbf{x}^{l} $ refers to the vector of \textbf{x} in layer $ l $.
The weights between neurons in a NN are represented as a weight matrix $  \textbf{W}$, where each element $ w_{ij} $ represents the weight between neuron $ i $ in the previous layer and neuron $ j $ in the current layer. 
A bias vector can be defined as an essential component that enables the model to fit data more effectively. 
The bias vector is a set of  constant values (one for each neuron) appended to the weighted input.
The bias term for each neuron is denoted as a vector $  \textbf{b}$, where each element $ b_i  $ indicates
the bias for neuron $ i $ in the current layer. 
It should be noted that the number of neurons may be chosen different for each layer. 

The output of a neuron in a layer can be calculated with the input vector dimension $ n $, as follows:

\begin{equation}
x^{(l)}_j= \sum_{i=1}^n(w^{(l)}_{ij}\,x_i^{(l-1)} + b^{(l)}_j).
\end{equation}

Then, the output of the neuron is passed through an activation function $ \sigma $ to produce the neuron of the next layer:

\begin{equation}
x^{(l-1)} \rightarrow x^{(l)}:=\sigma (W^{(l)}\,x^{(l-1)} + b^{(l)}).
\end{equation}

An activation function is a mathematical function that calculates the weighted sum of inputs and biases.
Selection of the activation function depend on the type of prediction problem and nature of the data.
It can be linear or nonlinear. Common activation functions include sigmoid: $ \sigma(x) = 1/(e^x +1) $, 
ReLU: $ \sigma(x) = max\{0,x\} $, softmax: $ \sigma(x_i) = e^x_i/(\sum_{j} e^x_j +1) $, 
and hyperbolic tangent: $  \sigma(x) = \tanh(x) $.

 Each layer in the network performs mathematical operations on 
 the input data, transforming it into a more useful representation for the next layer. 
 This process iterates over hidden layers, and 
continues until the output layer produces a final result.
During the training, weights and biases are updated.
In order to get the optimal set of weight and bias, we need to calculate
the error between the predicted output and the desired outcome and minimize it.
Thus, an optimization procedure is implemented to modify the weights and biases iteratively in order to reduce the error.
This process is called back-propagation which is a crucial step in training the NN.
A schematic of a neural network architecture including input, hidden, 
and output layers is illustrated in Fig. (\ref{DNN}).

\begin{figure}[h!]
 \begin{center} 
\includegraphics[width=0.7\textwidth]{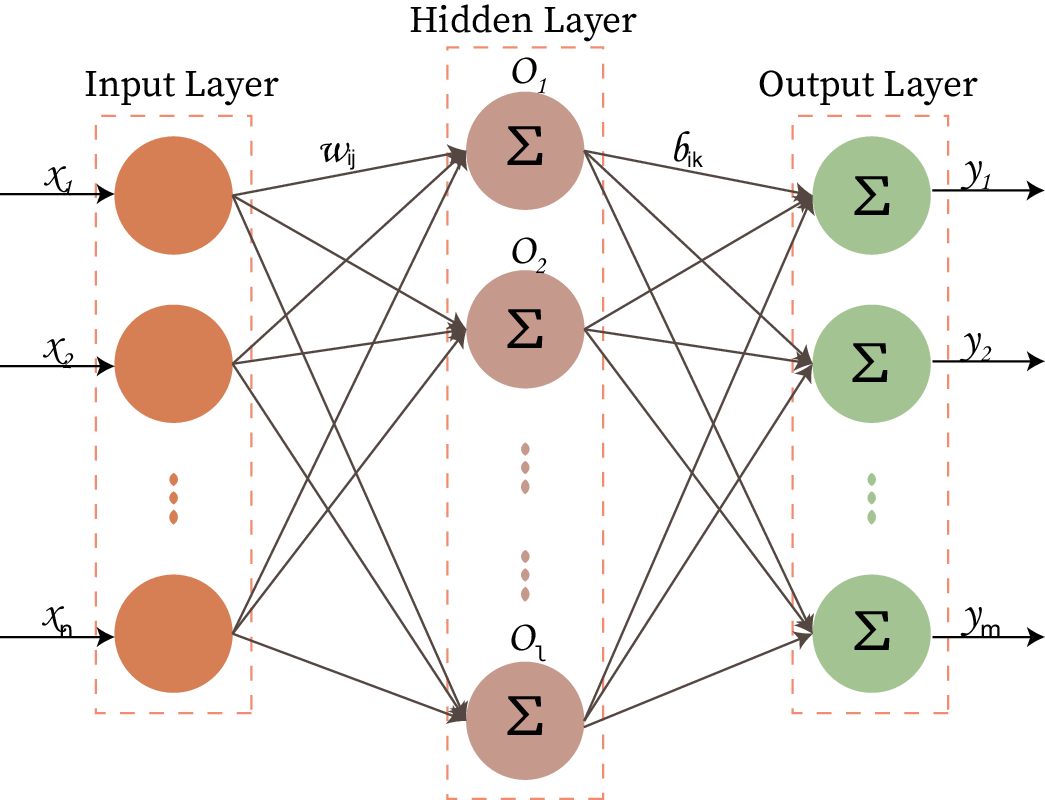}
\end{center}
\caption{Schematic illustration of a feedforward NN architecture.}\label{DNN}
\end{figure} 

Training a NN with a great performance to new (unseen) data can be a challenging problem. 
A model with too small and limited training data cannot learn the problem. 
In contrast, a model with too much capacity of training dataset can learn it too well and lead to overfitting. 
Specialized techniques are needed to avoid them.
The goal of a NN is to design a final model to fit the training data 
and make reliable predictions on the new data (test data). 

Validation is an effective technique in the ML to estimate performance of the model during learning. 
It is done by splitting the input dataset into the training and validating sets and 
then evaluating the predictive power of the model (DNN in this case) on the validation sets. 
In this study, the input data is randomly separated into two subsets 
as 90.0\% and 10.0\% for training and validation data respectively.

It is obvious that there is no universal rule for the NN architecture to get the best performance. 
Obtaining the appropriate number of neurons and the hidden layers as well as choosing the sufficient activation functions depend on the problem to be solved. 

In this work, to prevent the overfitting, and to train the DNN in a reasonable time, the number of hidden layers and nodes should not be too large. We found good performance considering the following points. 
In fact, number of the neurons on the hidden layers approximately equal to the average number of the input and output layers. Also, the number of hidden neurons keep on decreasing in subsequent layers.   

\subsection{Loss function and training the DNN:}
When we train a DNN, we feed the data to the network, make predictions, compare them with the actual output (the targets) and then calculate what is known as a loss. 
A loss function plays a key role to evaluate the performance of the NN model. 
It indicates how well the algorithm is learning the patterns in the training data.
The higher the loss the worse our model is performing.
Minimizing the loss function directly causes the model to make more accurate predictions.
The average, or expected  loss is then given by 

\begin{eqnarray}
\mathbb{E}[ {\cal L}] = \int \int   {\cal L}(\hat{y},y({\bf x})) p({\bf x},\hat{y}) d{\bf x} d\hat{y},
\end{eqnarray}
where  a specific estimate $ y({\bf x}) $ of the value of $ \hat{y}$ for each input $ {\bf x} $, and $ p({\bf x},\hat{y}) $ is a joint distribution over $ {\bf x} $ and  $ \hat{y}$. 
We are not required to know the exact form of the distribution. 
If we approximate it with the sum over the samples in the training dataset:

\begin{eqnarray}
\mathbb{E}[ {\cal L}] \approx \dfrac{1}{N}  \sum_{i=1}^N {\cal L}(\hat{y},y({\bf x})), 
\end{eqnarray}
where $N$ is the sample size.

As mentioned above, one way to determine the network parameters is
minimizing the loss function.  
Weights at the iteration step $ t $ are changed and updated along the direction of the derivative of the loss function with respect to the parameters:

\begin{eqnarray}
\theta^{(t+1)}_j  = \theta^{(t)}_j - \eta \dfrac{\partial{\cal L}^{(t)}}{\partial \theta_j }  \; \; \; \;\; \; \; \; \; \;with \; \;\; \; \; \; \; \;  \theta_j\in \{b,W\},
\end{eqnarray}
where $ \eta $ is the learning rate, and the minus sign implies that our optimization algorithm is navigating down the gradient.
Optimization algorithm is in charge of reducing the losses.
This naive form has some advantages and disadvantages. 
For example, this form is easy for computing, implementing, and understanding but may trap at local minima.
Weights and biases are changed and updated after calculating the gradient over the entire training dataset. 
Thus, if the dataset is too large, then it may take a lot of time to converge to the minima. 
%There are many different optimization algorithms that can be used to solve this problem. 
%Adam is one of the best and powerfull optimization algorithms for the DNNs.
%It's known for its effectiveness in training the DNNs, particularly in high dimensional parameter spaces. 
%We have examined different optimization algorithms to train the DNNs, and
%found the Adam as the best optimizer for our specific task.

Optimizer algorithms plays a crucial role in facilitating the training phase in a DNN model. 
In practice, optimizers are supposed to modify each epoch’s weights and reduce the loss function 
during the learning process. Different optimizers are used in the DL algorithms such as, Adam, SGD, AdaDelta and AdaGrad. 
The choice of an optimizer depends on many factors such as the model structure, size and configuration of the dataset. We implemented the Adam optimizer in our DNN model. 
In our study, as it is shown in Fig. (\ref{opti}) 
Adam yields better training loss and the NN model learns efficiently much better compared to the other optimizers.

\begin{figure}[h!]
 \begin{center} 
\includegraphics[width=0.7\textwidth]{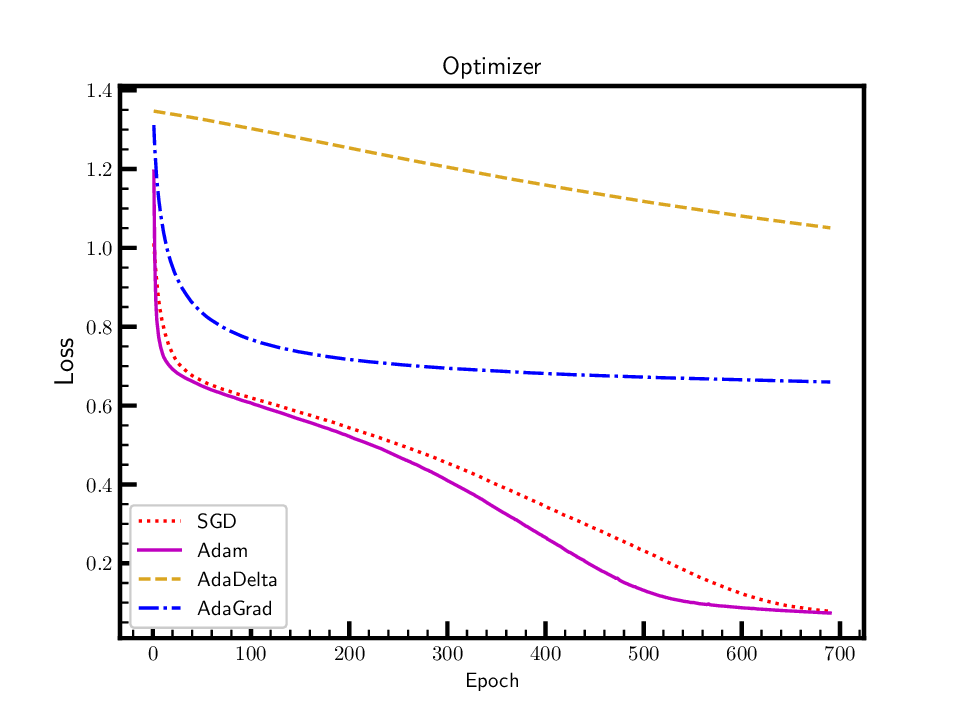}
\end{center}
\caption{The performance of Adam in the loss function in comparison with the other optimizers (SGD, AdaDelta and AdaGrad).}\label{opti}
\end{figure} 

An epoch can be explained as one complete passing of the training data through the algorithm.
The number of epochs is an important hyperparameter for the training process. 
We realized that a range between 500 to 1500 epochs is appropriate to balance between underfitting and overfitting in various analyses of this project. 

\begin{figure}[h!]
 \begin{center} 
\includegraphics[width=1.1\textwidth]{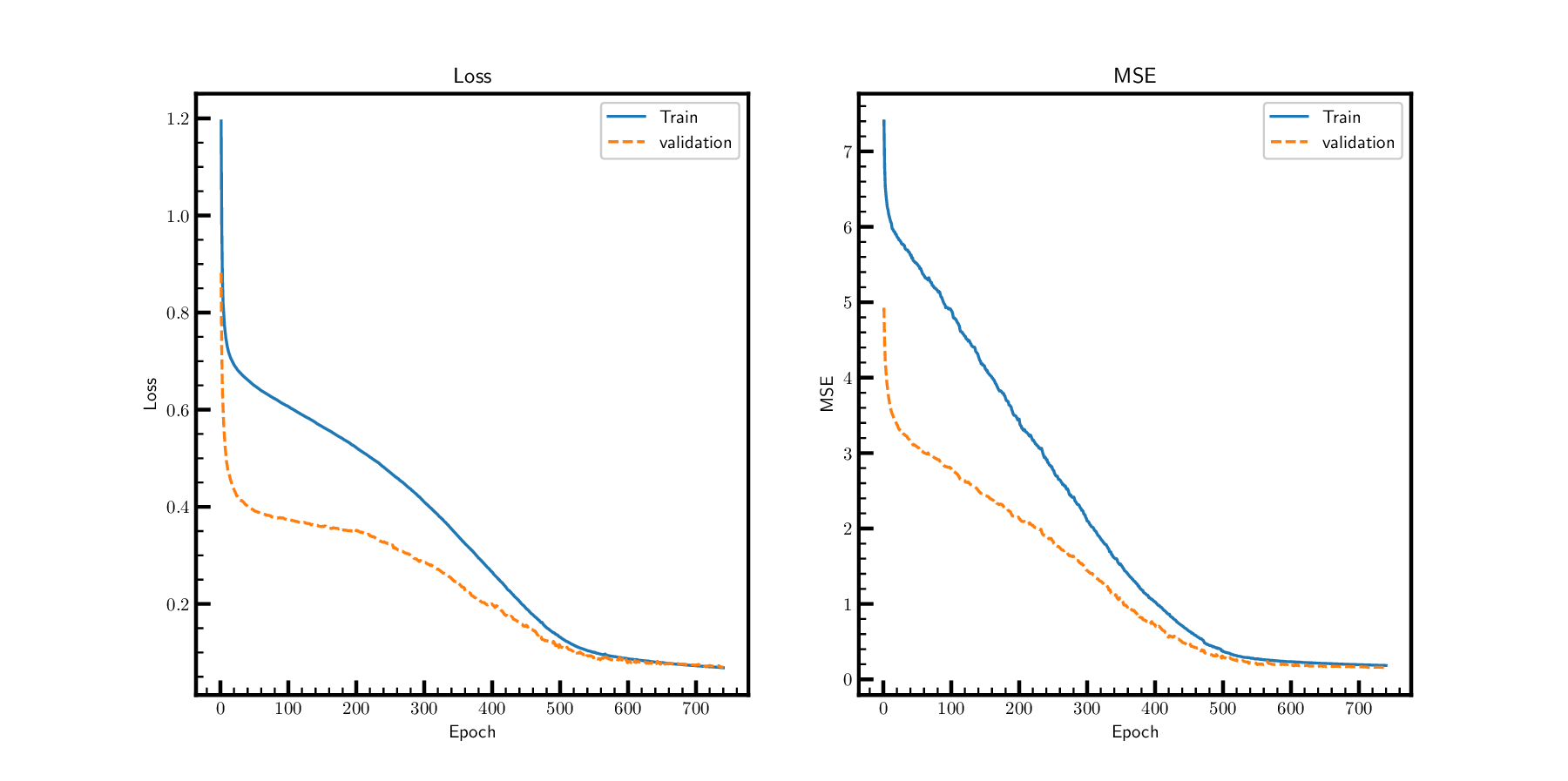}
\end{center}
\caption{The loss curves over the epochs for the training and validation datasets regarding the width prediction. The model presents comparable behaviour on the training and validation data.}\label{epoch}
\end{figure} 

Dropout method is generally used to reduce the overfitting of the NNs. 
Besides, batch normalization technique accelerates the training process and in some cases may improve the model performance. However, these methods were not efficient for our codes structure.

Generally, choosing the loss function is related to the activation function used in the output layer of the DNN. 
In other words, these two main functions are connected. 
For our setup, Tanh and Relu activation functions are considered in hidden layers and the linear activation function $ y = x $ for the output layer. Mean Squared Logarithmic Error (MSLE) and Mean  Squared  Error (MSE) are employed as the loss function and the metric respectively. 
MSLE  function measures the average of the squared differences between the logarithms of the predicted and actual values. It can be practical when the range of target values is large.
\begin{eqnarray}
\text{MSLE}(y, \hat{y}) = \frac{1}{N} \sum_{i=0}^{N - 1} (\log_e (1 + y_i) - \log_e (1 + \hat{y}_i) )^2,
\end{eqnarray}
where $ y_i $ is the actual value, $  \hat{y}_i $ is the predicted value, and $ N $ is the sample data size.
Also, Mean Squared Error (MSE) is one of the most commonly used metric for evaluation in ML models. 
MSE calculates the average squared difference between predicted and observed values. 
It is determined as the sum of the squared differences between the predictions and the observations, 
divided by the number of sample data. The formula for MSE is as follow,
  \begin{eqnarray}
\text{MSE}=  \frac{1}{n}\sum\limits_{i=1}^{n} (y_i - \hat{y}_i)^2.
\end{eqnarray}

Fig. (\ref{epoch}) as a sample, demonstrates the stability of the DNN model. It illustrates the behavior of the loss function on both training and validation datasets per epochs through a learning process, regarding the width prediction.
We have used a Python library, Keras \cite{Keras} with TensorFlow \cite{Abadi:2016kxi}.

\section{Data preprocessing strategy}\label{NA}
Data preparation for the ML techniques including NN, starts with data collection. 
It is a crucial step in every ML project, since it directly affects the model's performance and accuracy.
In this paper, the dataset has been collected from the PDG  \cite{ParticleDataGroup:2022pth}. 
Indeed, information of 376 mesons as well as exotic mesons has been precisely extracted. 
It must be emphasized that our dataset is more completed than those prepared in previous study \cite{Gal:2020dyc}.
At first approach, according to the Ref. \cite{Gal:2020dyc}, 
the dataset includes the quark content of mesons, their isospin ($I$), angular momentum ($ J $), and parity ($ P $) quantum numbers.  The variable $I$ can be $ 0, 1/2 $ and $ 1 $. $J$ varies from 0 through 6 in positive integers.
$ P $ only takes $ -1 $ or $ 1 $ values.
 It should be noted that, the mesons with the same quark structure and the identical quantum numbers but different masses may cause ambiguity in the NN performance. So, as to modify the input data and resolve ambiguities, another feature is defined as higher state ($ h $). Notice that $ h $ is not a real quantum number. 
This is a new parameter that just distinguishes the particles with the same properties and various masses.
For instance,$  f_0(980) $ and $ f_0(1370)  $ have the identical features in the input data and different masses. 
Consequently, they are given the $ h $ values of $ 0 $ and $ 1 $ respectively to be distinguishable as two separate entities for the DNN algorithm. The range of $ h $ can be changed from $ 0 $ to $ 10 $ depending to numbers of similar mesons. 
The value of $ h $ for mesons with unique properties is $ 0 $.
Furthermore, $ G $-parity, $ C $-parity as well as the mass and the decay width of the particles in units of MeV has been extracted. Finally, we have the following input vector,

\begin{eqnarray}
\label{eq}
\vec{v} = (d,\bar{d},u,\bar{u},s,\bar{s},c,\bar{c},b,\bar{b},I,J,P,h,G,C).
\end{eqnarray}

To determine the valence quarks of the mesons with linear combinations of pairs of quarks and antiquarks ($ q\bar{q} $), we have implemented conventions recommended in Ref.\cite{Gal:2020dyc}. For instance, the feature values of valence quarks of isotriplet mesons like $ \pi_0 $ and $ \rho_0 $ are $ (1, 1,  \vec{0}_8 , I, J, P, h, G, C) $.
The values of $ G $ or $ C $ will be zero if a meson is not a $ G $-parity or $ C $-parity eigenstate.
It is noteworthy that, the mass and the width values of the input mesons are scaled through common scaling techniques in data science to constrain the range of possible outputs. 
Since the mass and width parameters are dimensionful, it is divided by 1 MeV before scaling. 
Finally, the preparation of the dataset is completely done.

First and foremost, we are going to compute the mass of some ordinary and exotic mesons, 
 whose mass values have been confirmed by the PDG  \cite{ParticleDataGroup:2022pth}.
The results can demonstrate the strength of our designed DNNs in prediction of the mass.
The input dataset containing information of 341 mesons is randomly splitted 
into two subsets as 90.0\% and 10.0\% for the training and validation 
data respectively. As mentioned in section \ref{NN}, the performances during 
learning processes are monitored by the validation test. 
At the first step of calculation, the input data including thirteen feature columns, 
(the number of valence quarks,$  I, J, P $) is fed into the DNN. 
The DNN processes the data from the input node to the output node and returns the outputs (particles' mass) through deep learning algorithms. 
In the next step, to be more accurate,
 the variable $ h $ as 
the fourteenth feature, is added to the data and the deep learning algorithms are implemented again. 
Last step is allocated to examine the influence of adding the $ G $ and $ C $ parities to the data features. 
The numerical results are demonstrated in tables \eqref{tab:wellmassA1} to  \eqref{tab:newexr}.
Our predictions based on the first approach, for the mass along with a comparison to the experiment  \cite{ParticleDataGroup:2022pth} and the other NN estimation \cite{Gal:2020dyc} 
are illustrated in tables \eqref{tab:wellmassA1},  \eqref{tab:lightmassA1}, and \eqref{tab:exomassA1}. 
Interestingly, the tables \eqref{tab:wellmassA1} and \eqref{tab:wellmassA2} draw attention to the performance of the DNN on predicting the mass of conventional mesons vs. corresponding tetraquarks. 
Section \ref{r} interprets the results in details.
Accordingly, we aim to predict the width of the normal and exotic mesons whose widths have
not been confirmed by the PDG. For this purpose, the mass of particles is appended to the feature columns in the dataset and the same procedure mentioned above, is performed. 
Our results for the width predictions, based on the first approach, are given 
in tables  \eqref{tab:widthA1} and \eqref{tab:widthA12},
and discussed in section \ref{r}.

The architecture of the dataset indicates how the information is accessed and organized. 
Namely, designing an efficient and appropriate data structure, significantly 
improves the DNN algorithm’s performance. 
In the first approach, the data structure was categorized according to the way proposed in Ref. \cite{Gal:2020dyc}.
In fact, first ten columns of the dataset belong to the number of valence quarks of each meson. 
Although, this procedure of organizing the data, looks to be straightforward, 
 works well just for mesons with $ q\bar{q} $ structure.  
There exist troubles in specifying the number of valence quarks of the mesons with linear 
combinations of $ q\bar{q} $ pairs in the dataset. 
 For instance, the flavor wave functions of the mesons of $ \pi_0 $ and $\eta  $ are,

 \begin{eqnarray}
\pi_0 & = &\dfrac{(u\bar{u} -d\bar{d})}{\sqrt{2}},\nonumber\\
\eta &= &\dfrac{(u\bar{u} +d\bar{d}- 2 s\bar{s})}{\sqrt{6}}.
\end{eqnarray}

As mentioned before, in the first approach the procedure recommended  in the Ref. \cite{Gal:2020dyc} were applied.
For instance, the isotriplet mesons such as  $\pi_0$ are encoded as $  (1,1,\vec{0}_8,I,J,P)  $, and $\eta$ which  is included in the category of the lighter isosinglet mesons, encoded as $  (\vec{0}_2,1,1,\vec{0}_6,I,J,P)  $.
Although this approach is not exactly describing the quark content of all kinds of mesons, 
the DNNs have proven to be strong enough to predict the mass and width of particles with a 
fair approximation. To protect the credibility and reliability of the data as well as enhancing the accuracy, 
we propose a novel idea for preprocessing the dataset. In this procedure, the quark content of mesons 
in association with the values of corresponding Clebsch-Gordan coefficients will be considered.  
So the first twenty five entries of the dataset are devoted to possible $ q\bar{q} $ states 
constructing the mesons' structure, 
like $ d\bar{d} $, $ d\bar{u} $, $ c\bar{d} $, and etc. 
Consequently, every meson with linear combinations of pairs of quarks
 and antiquarks, gets the corresponding Clebsch-Gordan coefficient of $q\bar{q}$ state as the feature value,
 unlike the previous approach that the number of valence quarks of a given type were 
 counted. The rest columns of the dataset is not changed and will be the same as the first approach. 
 Consequently, the dataset is optimized and becomes a more effective framework. 
 For example, the input vector of $\eta$, ($I^G(J^{PC}) = 0^+(0^{-+})$)  considering all features $  (\overrightarrow{q\bar{q}}_{25}, I, J, P, h, C, G)  $ has the following form,

 \begin{eqnarray}
\eta = (\dfrac{1}{\sqrt{6}},\vec{0}_4,\dfrac{1}{\sqrt{6}},\vec{0}_6,\dfrac{-2}{\sqrt{6}},\vec{0}_{12},0,0,-1,0,1,1).
\end{eqnarray}

Now, the modified dataset is ready for analyses. 
Same as the previous approach, the DNN evaluates the input data to 
make predictions on the test data in three levels. 
Firstly, the input data contains the Clebsch-Gordan coefficient of $ q\bar{q} $ states 
of the mesons in association with the $ I, J, P $ as the features. In the second level, the higher state 
of $ h $ is appended to the data features. 
Thirdly, the $ C $ and $ G $ variables are considered too. 
Our predictions based on the second approach, for the mass compared to the experiment  \cite{ParticleDataGroup:2022pth} and the other NN estimation \cite{Gal:2020dyc} 
are illustrated in tables \eqref{tab:wellmassA2},  \eqref{tab:lightmassA2}, and \eqref{tab:exomassA2}. 
It should be emphasized that the main network 
architecture, including the number of layers, scaling methods, 
setting the optimizer, loss functions and metric remains unchanged. 
Certainly, increasing the nodes of each layer is inevitable due to the 
enhancing the features.  Besides, by adding the mass as another 
feature to the dataset, we repeat this procedure to predict the mesons
 width values. For instance, the details of our designed DNN for 
 prediction of the mass considering all the sixteen features based 
 on the first approach (A1) is summarized in table \eqref{tab:layer}. Practically, 
 similar DNN configurations are proposed for other predictions 
 of the mass and width based on the A1 and A2. Note that, size 
 of the input layer is equal to the number of feature columns.
All the numerical results are disscused in section \ref{r} completely.
 
\begin{table}[h]
\begin{center}
\scalebox{1.0}{\begin{tabular}{l*{3}{c}r}
\hline
Layer (type) & Number of neurons &Activation function \\
\hline
Dense &16& tanh\\
Dense &12 &tanh\\
Dense &8 &tanh\\
Dense &4 &relu\\
Dense &1 &linear\\
\hline
\end{tabular}}
\caption{The DNN architecture designed for prediction of the mass considering all the sixteen features based on the first approach (A1).} \label{tab:layer}
\end{center}
\end{table}

 \section{ Interpretations of the results}\label{r}
In this section, we present and discuss our numerical results determined by the DNNs. 
As pointed out above, we adopt two distinct approaches to construct the dataset. 
In the both approaches, the data features are being completed in three steps.
So the data is fed into the DNN with three sets of input vectors.
For prediction of the mass, at the first step, the data are classified based on
the mesons' quark content and their quantum numbers $  I $,  $J$, and $P$ (base data).
Next, the variable $h$ is appended to the base data (base data  $+~h$).
At last, the quantum numbers  $G$ and $C$ are added to the features (base data  $+~h\,G\,C$). 
Important to underline that, in the first approach (A1), 
the number of valence quarks of each meson are entered in the first ten columns of the dataset.
Therefore, the input vector (\ref{eq}) denotes the most complete version of the data features.
In contrast, the second approach (A2) considers the Clebsch-Gordan coefficients of all possible $q\bar{q}$ structures in the quark content of a typical meson, leading to a 25-dimensional vector in the dataset.
Correspondingly,  for prediction of the width, the mass of particles is appended to the features 
in the dataset and the same procedure mentioned above, are implemented. 

We aim to predict the mass of some well-known, light and exotic mesons, 
as well as the width of mesons that have not been experimentally determined. 
The numerical results for prediction of the mass and decay width
as well as the corresponding mean errors 
are demonstrated in tables \eqref{tab:wellmassA1} to \eqref{tab:newexr}.
The tables \eqref{tab:wellmassA1} to \eqref{tab:exor} show the predicted mass and 
  the tables \eqref{tab:widthA1} to \eqref{tab:widthA22} illustrate the estimated width results.
 We also estimate the mass and width of four new tetraquarks, through the DNNs, summarized
  in tables \eqref{tab:newtetA1} to \eqref{tab:newexr}.

\subsection{$q\bar{q}$ \textit{versus} $q\bar{q}q\bar{q}$}
The structure of some famous mesons have been actively discussing for years. 
The question is whether the $a_0(980)$, $f_0(980)$, $D^*_{s0}(2317)^{\pm}$ and $D_{s1}(2460)$
are grouped in the ordinary mesons or they belong to the exotic states. 
We intend to predict the mass of these challenging mesons using the DNN, based on two assumptions. 
We first suppose the $q\bar{q}$ state and then examine the $q\bar{q}q\bar{q}$ state for their quark contents.
Tables \eqref{tab:wellmassA1} and \eqref{tab:wellmassA2} 
show the numerical results considering both quark contents, obtained 
according to A1 and A2 respectively. 
The DNNs are supposed to predict the mass considering 
three sets of dataset (base data, base data  $+~h$ and base data  $+~h\,G\,C$).
As previously stated, the data features become more complete step by step.
Entirely, the predicted results for the mass, assuming two different 
structures, based on two approaches,
are found to be comparable with the experimental data.
Let's compare the obtained results according to the quark content, the approach and the dataset features.
We calculated the mean error of the predicted masses with the observed ones too, shown in table \eqref{tab:wellr}. One can observe that the mean error  with respect to the observed results for the $q\bar{q}$  structure is much lower than the tetraquark $q\bar{q}q\bar{q}$
one, suggesting the $q\bar{q}$ structure for the corresponding mesons.

In the case of the $a_0(980)$ meson, pursuant to A1,
the DNN predictions for the mass, are getting close to the experimental value 
($980\pm 20$ MeV) with completing the data features. 
More precisely, when the $q\bar{q}$ configuration is considered,
the DNN, forecasts a mass of $998\pm {94}$ MeV by importing the base data $ +~h\,G\,C $. 
While the $a_0(980)$ with $q\bar{q}q\bar{q}$  structure is given to the DNN, 
the mass is relatively well estimated at $1069\pm {224}$ MeV, 
in the most complete version of the data.
Similarly, if we notice the results based on A2,
the DNN predictions for the $a_0(980)$ mass, are improving step by step with enhancing the data features.
Supposing the $q\bar{q}$ picture, the predicted mass is $989\pm 93$ MeV, 
and considering $a_0(980)$ as the tetraquark, 
the mass is obtained $1152\pm97 $ MeV, in the most complete version of the data features.
These observations highlight critical importance of the model selection 
in theoretical predictions and underscore the need for empirical validation to refine these models further.
According to the PDG \cite{ParticleDataGroup:2022pth}, the mass of the $f_0(980)$ is  $990\pm20 $ MeV.
Whenever the $f_0(980)$ is supposed to be an ordinary meson,
the predicted mass via our DNN is more compatible with the experimental data, per the A1 and A2.
Best prediction for the mass value is obtained $966\pm51$ MeV, 
when the base data $+~h $, according to the A1, is fed into the DNN. 
This assertion holds true for the mesons $D_{s_0}^{*}(2317)^\pm$ and $D_{s_1}(2460)^\pm$.
Our DNNs mass predictions exhibit diminished discrepancies when juxtaposed with experimental measurements, surpassing the precision of the estimates presented in Ref. \cite{Gal:2020dyc}. 
This confluence of accuracy across multiple mesonic cases not only bolsters the credibility of our DNNs performance, but also provides compelling evidence 
for its superiority in capturing the nuances of mesonic mass spectra.

Furthermore, incorporating the parameter $h$ into the dataset (base data  $+~h$) markedly reduces the mean error more substantially than the base data alone, while the inclusion of $C$ and $G$ quantum numbers does not enhance the model's accuracy. The minimal mean error achieved is $1.74 \%$, realized through the A1 within the base data  $+h$ dataset. 
The A2 also serves as a competent predictor, yields a mean error of $3.49 \%$. 
These results underscore the efficacy of integrating higher states 
into the dataset for precision modeling in particle physics.

Lastly, we find that, our DNN predictions for the mass of these controversial mesons are more compatible to 
the experimental ones, when they are classed as the ordinary $q\bar{q}$ mesons for both approaches.
Nevertheless, the actual quark content of these mesons still has remained ambiguous.

 \begin{table}[h!]
\begin{center}

\renewcommand{\arraystretch}{1.4}
\scalebox{0.66}{
\begin{tabular}{|c|c|c|c|c|c|c|c|}\cline{1-8}
 Meson&  $I^G\,(J^{PC})$	 &   Exp. Mass (MeV)  \cite{ParticleDataGroup:2022pth}  &  quark content & Ref.\cite{Gal:2020dyc} &base data  &  base data $ +~h $   &  base data $ +~h\,G\,C  $ \\ \hline
 \hline
\multirow{2}{*}{$a_0(980)$}	 &\multirow{2}{*}{ $1^-\,(0^{++})$}	& \multirow{2}{*}{$980\pm 20$} & $u\bar{u}$    &   $1277\pm 246$ &   $1312\pm {54}$   &  $1020\pm {69}$&  $998\pm {94}$\\ \cline{4-8}
& & & $u\bar{s}\bar{u}s$ $(K\overline{K})$   &   $2172\pm 466$  & $1340\pm {104}$   &  $1100\pm {104}$ & $1069\pm {224}$ \\ \hline
\multirow{2}{*}{$f_0(980)$}	 &\multirow{2}{*}{ $0^+\,(0^{++})$}	& \multirow{2}{*}{$990\pm 20$} & $d\bar{d}$   &   $921\pm 117$ &    $1457\pm {63}$  &  $966\pm {51}$  &  $883\pm {45}$  \\ \cline{4-8}
& & & $d\bar{s}\bar{d}s$ $(K\overline{K})$   &   $1592\pm 401$   &  $1723\pm {125}$ &  $1155\pm {104}$& $1086\pm {68}$\\ \hline
\hline
\multirow{2}{*}{$D_{s0}^*(2317)^\pm$}	 &\multirow{2}{*}{ $0\,(0^+)$}	& \multirow{2}{*}{$2317.8\pm0.5$} & $c\bar{s}$   & $2640\pm 433$&  $2289\pm 183$  & $2322\pm {142}$ & $2343\pm {169}$ \\ \cline{4-7}\cline{8-8}
& & & $c\bar{u}u\bar{s}$ $(DK)$   &   $4326\pm 925$   & $3091\pm {361}$ &  $2471\pm {284}$& $2511\pm {334}$\\ \hline
\multirow{2}{*}{$D_{s1}(2460)^\pm$}	 &\multirow{2}{*}{ $0\,(1^+)$}	& \multirow{2}{*}{$2459.5\pm 0.6$} & $c\bar{s}$    &   $2547\pm 39$   & $2356\pm {121}$ & $2453\pm {128}$&$2442\pm {218}$\\ \cline{4-7}\cline{8-8}
& & & $c\bar{u}u\bar{s}$ $(D^*K)$   &   $3431\pm 544$  &  $2845\pm {251}$  & $2527\pm {289}$&$2748\pm {504}$\\ \hline
\end{tabular}
}
	\caption{Our DNN predictions for the mass of four well-known conventional mesons vs. corresponding tetraquark structures (in units of MeV), based on the A1, 
in comparison with the experimental \cite{ParticleDataGroup:2022pth} and 
the other NN results in Ref.\cite{Gal:2020dyc}.}\label{tab:wellmassA1}
 \end{center}
\end{table}

\begin{table}[h!]
\begin{center}
\renewcommand{\arraystretch}{1.4}
\scalebox{0.66}{
\begin{tabular}{|c|c|c|c|c|c|c|c|}\cline{1-8}
Meson&  $I^G\,(J^{PC})$	 &  Exp. Mass (MeV)\cite{ParticleDataGroup:2022pth} &  quark content & Ref.\cite{Gal:2020dyc} & base data  &  base data $ +~h $   &  base data $ +~h\,G\,C  $ \\ \hline
\hline
\multirow{2}{*}{$a_0(980)$}	 &\multirow{2}{*}{ $1^-\,(0^{++})$}	& \multirow{2}{*}{$980\pm 20$} & $u\bar{u}$    &   $1277\pm 246$ &   $1079\pm 91 $   &  $987\pm 12$&  $989\pm93 $\\ \cline{4-8}
& & & $u\bar{s}\bar{u}s$ $(K\overline{K})$   &   $2172\pm 466$   &   $1490\pm108 $   &  $1041\pm200$&  $1152\pm97 $ \\ \hline
\multirow{2}{*}{$f_0(980)$}	 &\multirow{2}{*}{ $0^+\,(0^{++})$}	& \multirow{2}{*}{$990\pm 20$} & $d\bar{d}$   &   $921\pm 117$  &      $1793\pm276 $   &  $1051\pm175$&  $1111\pm40 $\\ \cline{4-8}
& & & $d\bar{s}\bar{d}s$ $(K\overline{K})$   &   $1592\pm 401$    &   $1792\pm250 $   &  $1251\pm226$&  $1216\pm216 $\\ \hline
\hline
\multirow{2}{*}{$D_{s0}^*(2317)^\pm$}	 &\multirow{2}{*}{ $0\,(0^+)$}	& \multirow{2}{*}{$2317.8\pm0.5$} & $c\bar{s}$   & $2640\pm 433$ &    $2975\pm79 $   &  $2456\pm107$&  $2308\pm 112 $ \\ \cline{4-7}\cline{8-8}
& & & $c\bar{u}u\bar{s}$ $(DK)$   &   $4326\pm 925$    &   $3997\pm314 $   &  $3912\pm184$&  $3819\pm 210 $\\ \hline
\multirow{2}{*}{$D_{s1}(2460)^\pm$}	 &\multirow{2}{*}{ $0\,(1^+)$}	& \multirow{2}{*}{$2459.5\pm 0.6$} & $c\bar{s}$    &   $2547\pm 39$    &    $2537\pm 84$   &  $2432\pm87$&  $2490\pm81 $\\ \cline{4-7}\cline{8-8}
& & & $c\bar{u}u\bar{s}$ $(D^*K)$   &   $3431\pm 544$   &   $3779\pm250$   &  $3466\pm274$&  $3386\pm253 $\\ \hline

\end{tabular}
}
	\caption{Our DNN predictions for the mass of four well-known conventional mesons vs. corresponding tetraquarks structures (in units of MeV), based on the A2, 
in comparison with the experimental \cite{ParticleDataGroup:2022pth} and 
the other NN results in Ref.\cite{Gal:2020dyc}.}\label{tab:wellmassA2}
\end{center}
\end{table}  
 
 \begin{table}[h!]
	\begin{center}
		\renewcommand{\arraystretch}{1.4}
		\scalebox{0.68}{
			\begin{tabular}{|c|c|c|c||c|c|c|}\cline{1-7}
				$ q\bar{q} $  vs $ q\bar{q}q\bar{q} $&  $ q\bar{q} $ A1\%	 &$ q\bar{q} $ A2\% & $ q\bar{q} $   Ref.\cite{Gal:2020dyc} \%&   $ q\bar{q} q\bar{q}$ A1\%	 &$ q\bar{q}q\bar{q} $ A2\%  & $ q\bar{q}q\bar{q}  $   Ref.\cite{Gal:2020dyc} \%  \\ \hline
				\hline  
				base$  $ & 21.63 & 30.68 &\multirow{3}{*}{13.68}	 & 39.95 & 64.78& \multirow{3}{*}{ 77.15}	\\   \cline{1-3} \cline{5-6}
				base$ +\, h $ & 1.74 & 3.44 & & 28.70 & 35.57&\\  \cline{1-3} \cline{5-6}
				base$ +\, h \,G\,C$ &3.61& 3.70 & & 9.71 & 35.70&\\  
				\hline
			\end{tabular}
		}
		\caption{ The mean errors of the predicted mass of four well-known mesons
		based on the A1 and A2, in comparison with  Ref.\cite{Gal:2020dyc}.}  \label{tab:wellr}
	\end{center}
\end{table} 
 
 We considered the error estimate based on the $n$ subset cross-validation technique. This technique splits the training data into $n$ subsets, and trains the NN algorithm, n times, each time using $n-1$ subsets for training and the remaining subset for validation. 
So we can calculate the error (e.g., mean squared error) on the validation set and the average error across all n subsets to estimate the model's performance. The final average error obtained from the $n$ subset cross-validation process can be used to calculate the model's generalization error on new, unseen data.

\subsection{\textit{Light mesons}}
We present predictions for the mass of some light mesons with $q\bar{q}$ structure 
per A1 and A2, in tables \eqref{tab:lightmassA1} and \eqref{tab:lightmassA2} respectively,
which have not been predicted in Ref. \cite{Gal:2020dyc}.
The mass of the $f_0(500)$ meson has not been exactly determined and 
the PDG \cite{ParticleDataGroup:2022pth} only gives a range of possible values. 
Firstly, we feed the dataset prepared accordance with the A1 into the DNN.
Using only the base data, the DNNs prediction falls outside the experimental range. 
However, when we include the effects of the  higher state $h$ and the quantum numbers $C$ and $G$,  our prediction agrees with the PDG range. 
This shows the importance of taking into account these symmetries in the mass spectrum of light mesons \cite{Brambilla:2019esw,Close:2002zu}. 
Subsequently, when the data features are modified based on the A2,
considering each set of dataset (base data, base data $+~h$ and base data $+~h\,G\,C$),
the DNNs exhibit remarkable performance in estimating the mass of the $f_0(500)$.

For the $K_4(2500)$ meson, predictions based solely on the base data per
two approaches, are close and slightly smaller than the expected value, considering the range of uncertainties.  Once the data features are optimized based on the A2,
considering the base data $+~h$ or the base data $+~h\,G\,C$,
the predictions coincide fairly well with the observed mass values within the uncertainty range.
 In fact, involving the Clebsch-Gordan coefficients in the data features,
 yielding more consistent results across the experimental values.

In 1979, the $ K_2(1580) $ meson was detected via a partial-wave analysis of the $ K^- \pi^+ \pi^- $ system. The mass spectrum of this meson was empirically approximated to be in the vicinity of $1580$ MeV Ref. \cite{Aachen-Berlin-CERN-London-Vienna:1978zeh}. However, this estimation has yet to receive confirmation from the PDG \cite{ParticleDataGroup:2022pth}.
Our DNNs provide a reasonable estimation for the mass of the $K_2(1580)$ meson as well. Employing the modified dataset based on the A2, which includes the base data $+~h\,G\,C$, the DNNs forecast a mass of $1634 \pm 52$ MeV for the $K_2(1580)$ meson. This prediction align closely with the experimental findings within the margin of uncertainty. Moreover, alternative mass estimations for the $K_2(1580)$, particularly those based on the base data within the A2, are in concordance with theoretical projections. Ref. \cite{Taboada-Nieto:2022igy} notably cites a mass of approximately  $1.75$ GeV for this meson.

We calculate the mean errors of the predicted mass of these light mesons with $q\bar{q}$ structure 
relative to their experimental values. 
For the $f_0(500)$, we use mean of  the mass range as the input.
The results are shown in table \eqref{tab:lightr}. 
We observe that, the contribution of the Clebsch-Gordan coefficients in the quark content, 
gives a better agreement with the experimental data than the A1. 
This demonstrates the significance of the Clebsch-Gordan coefficients in the mass formula of light mesons. 
The best fit is obtained when we also consider the higher state number $h$ and
 the quantum numbers $C$ and $G$. 
 In this case, the mean error of the masses is reduced to $4.14 \%$.

 \begin{table}[h!]
\begin{center}

\renewcommand{\arraystretch}{1.4}
\scalebox{0.68}{
\begin{tabular}{|c|c|c|c|c|c|c|}\cline{1-7}
 Meson&  $I^G\,(J^{PC})$	 &  Exp. Mass (MeV)  \cite{ParticleDataGroup:2022pth}  &  quark content  &base data  &  base data $ +\, h $   &  base data $ +\, h\,G\,C $  \\ \hline
\hline
$f_0(500)$ & $0^+\,(0^{++})$ & $400-800$  & $d\bar{d}$   & $1748\pm {16}$ & $668\pm {61}$ &$759\pm {134}$ \\ \hline

$K_{4}(2500)_0,\bar{K_{4}}(2500)_0$ & $1/2\,( 4^-)$ & $2490\pm {20}$  & $d\bar{s}, s\bar{d}$   & $2232\pm {60}$ & $2278\pm {57}$ &$2308\pm {35}$ \\ \hline

$K_{4}^{\pm}(2500)$ & $1/2\,( 4^-)$ & $2490\pm {20}$  & $u\bar{s}, s\bar{u}$ & $2206\pm {28}$ & $2283\pm {51}$ &$2298\pm {25}$ \\ \hline

$K_{2}(1580)_0,\bar{K_{2}}(1580)_0$ & $1/2\,( 2^-)$ & $1580$  & $d\bar{s}, s\bar{d}$   & $1787\pm {31}$ & $1645\pm {25}$ &$1646\pm {20}$ \\ \hline

$K_{2}^{\pm}(1580)$ & $1/2\,( 2^-)$ & $1580$  & $u\bar{s}$  & $1661\pm {28}$ & $1687\pm {31}$ &$1653\pm {23}$ \\ \hline

\end{tabular}
}	\caption{Our DNN predictions for the mass of some light mesons (in units of MeV), based on the A1, 
in comparison with the experimental \cite{ParticleDataGroup:2022pth} and 
the other NN results in Ref.\cite{Gal:2020dyc}.}\label{tab:lightmassA1}
 \end{center}
\end{table}

\begin{table}[h!]
\begin{center}
\renewcommand{\arraystretch}{1.4}
\scalebox{0.68}{
\begin{tabular}{|c|c|c|c|c|c|c|}\cline{1-7}
Meson&  $I^G\,(J^{PC})$	 &  Exp. Mass (MeV) \cite{ParticleDataGroup:2022pth} & quark content&base data  &  base data $ +\, h $   &  base data $ +\, h\,G\,C $   \\ \hline
\hline
$f_0(500)$ & $0^+\,(0^{++})$ & $400-800$  & $d\bar{d}$   &   $754\pm278 $   &  $448\pm285$&  $554\pm128 $ \\ \hline

$K_{4}(2500)_0,\bar{K_{4}}(2500)_0$ & $1/2\,( 4^-)$ & $2490\pm {20}$  & $d\bar{s}, s\bar{d}$  & $2185\pm125$ & $2445\pm135$ &$2510\pm112$ \\ \hline

$K_{4}^{\pm}(2500)$ & $1/2\,( 4^-)$ & $2490\pm {20}$  & $u\bar{s}, s\bar{u}$   &   $2286\pm114 $   &  $2459\pm135$&  $2558\pm128 $ \\ \hline

$K_{2}(1580)_0,\bar{K_{2}}(1580)_0$ & $1/2\,( 2^-)$ & $1580$  & $d\bar{s}, s\bar{d}$  &   $1754\pm78 $   &  $1696\pm66$&  $1676\pm57 $ \\ \hline

$K_{2}^{\pm}(1580)$ & $1/2\,( 2^-)$ & $1580$  & $u\bar{s}$   &   $1789\pm83 $   &  $1662\pm42$&  $1634\pm52 $ \\ \hline

\end{tabular}
}
	\caption{Our DNN predictions for the mass of some light mesons (in units of MeV), based on the A2, 
in comparison with the experimental \cite{ParticleDataGroup:2022pth} and 
the other NN results in Ref.\cite{Gal:2020dyc}.}\label{tab:lightmassA2}
\end{center}
\end{table} 
 
\begin{table}[h!]
	\begin{center}
		\renewcommand{\arraystretch}{1.4}
		\scalebox{0.68}{
			\begin{tabular}{|c|c|c|}\cline{1-3}
				light meson& A1\%	 & A2\%  \\ \hline
				\hline  
				base$  $ &46.27&14.07\\  \hline
				base$ +\, h $ & 7.81&8.18 \\  \hline
				base$ +\, h \,G\,C$ &10.06&4.14\\  
				\hline
			\end{tabular}
		}
		\caption{The mean errors of the predicted masses of the light mesons,
		based on the A1 and A2.}  \label{tab:lightr}
	\end{center}
\end{table} 

\subsection{\textit{Exotic Mesons}}
The DNN are trained to predict the mass of some exotic mesons with $q\bar{q}q\bar{q}$ structure, 
whose masses are confirmed by the PDG \cite{ParticleDataGroup:2022pth}. 
The DNN predictions regarding the A1 and A2 are presented 
in tables \eqref{tab:exomassA1} and \eqref{tab:exomassA2} respectively. 
We compare the DNN output values with the experimental ones and 
calculate the mean errors, which are shown in table \eqref{tab:exor}. 
We find that our deep learning algorithms outperforms 
the previous model of Ref. \cite{Gal:2020dyc} in terms of accuracy. 
We also observe that the A2 gives better results than the A1. 
This implies the significance of considering the Clebsch-Gordan coefficients in the quark content of ordinary mesons, which enhance the performance of the model even for the prediction of mass values of exotic states.
The best fit will be achieved, when we include the effects of the higher state number $h$ and the quantum numbers $C$ and $G$. In this case, the mean error of the masses is reduced to $16.01 \%$. 

Overall, our DNNs exhibit a higher precision in predicting the mass of the ordinary mesons  
relative to the exotics, compared to the current experimental data, as evidenced by the errors detailed 
in tables \eqref{tab:wellr}, \eqref{tab:lightr} and \eqref{tab:exor}. 
This outcome is anticipated, given that our training dataset is exclusively composed of ordinary mesons. 

The enhancement of DNN performance through the integration of internal structures and quantum numbers into the dataset is a testament to the intricate relationship between detailed physical attributes and machine learning efficacy. By embedding these fundamental aspects, our model can achieve a more nuanced understanding and prediction, leading to more accurate and insightful outcomes.
For instance, in the case of the $\psi(4660)$ meson, which has an experimentally observed mass of $(4630 \pm 6)~ \mathrm{MeV}$, our initial approach (A1) using only the base data predicts a mass of $(3442 \pm 137)~ \mathrm{MeV}$, deviating from the observed value. However, by incorporating the higher state $h$ into the dataset based on the A2, the predicted mass improves to $(4186 \pm 671)~ \mathrm{MeV}$, aligning within the experimental uncertainty range. 
Once more, this underscores the importance of involving the Clebsch-Gordan coefficients 
in the data features as well as contribution of higher state $h$ in raising 
the accuracy of mass predictions for exotic mesons.

In the analysis of certain exotic mesons such as $R_{c0}(4240)$,
 importing the base data $ +~h\,G\,C $, based on the A1,
yields predictions that concur with the observed mass values within their uncertainty bounds.
 Conversely, the mass predictions for $R_{c0}(4240)$, based on the A2, lied lower than the experimental ones, 
 despite the inclusion of the higher state $h$ and quantum numbers $C$ and $G$. 
 This discrepancy indeed emphasizes the intricate challenges inherent in modeling exotic mesons. 
 It raises the intriguing possibility that alternative structures, such as sexaquarks, may be more appropriate assignments for these mesons. Consequently, there is a pressing need to refine the training set with data specific to exotic mesons. Such targeted data enrichment would enable the models to more accurately encapsulate and reflect the unique properties of these mesons, thereby enhancing the fidelity of theoretical predictions within the quantum chromodynamics framework.

 \begin{table}[h!]
\begin{center}
\renewcommand{\arraystretch}{1.4}
\scalebox{0.59}{
\begin{tabular}{|c|c|c|c|c|c|c|c|}\cline{1-8}
 Meson&  $I^G\,(J^{PC})$	 &  Exp. Mass (MeV) \cite{ParticleDataGroup:2022pth}  & quark content & Ref.\cite{Gal:2020dyc} &base  data &  base data $ +\, h $   &  base data $ +\, h\,G\,C $   \\ \hline
 \hline
$\chi_{c1}(3872)$	 
& $0^+\,(1^{++})$	& $3871.65\pm 0.06$  & $c\bar{u}\bar{c}u$ $(D^0\,\bar{D}^{*0})$ &   $4815\pm 786$ & $3309\pm {174}$  &$2949\pm {159}$ &$2944\pm {177}$ \\ \hline
$\psi(4230)$ & $ 0^-\,( 1^{--})$ & $4222.5\pm 2.4$  & $c\bar{s}\bar{c}s$ $(D_s\,\bar{D}_s)$ &  $(5.4\pm 1.1)\times 10^3$  & $3200\pm {176}$ & $3358\pm {246}$ &$3303\pm {175}$ \\ \hline
$\psi(4360)$ & $ 0^-\,( 1^{--})$ &  $4374\pm 7$  &  $c\bar{u}\bar{c}u$ $(D_1\,\bar{D}^*)$ & \multirow{2}{*}{$4940\pm 903$} & \multirow{2}{*}{ $3442\pm {137}$} & $3422\pm {240}$& $3190\pm {184}$ \\ \cline{1-4}
$\psi(4660)$ & $0^-\,( 1^{--})$ & $4630\pm 6$  &  $c\bar{u}\bar{c}u$ $(f_0(980)\,\psi^\prime)$ &  & &  $3480\pm {253}$&  $3300\pm {79}$ \\ \hline
$Z_c(3900)^\pm$ & $ 1^+\,( 1^{+-})$	&$3887.1\pm 2.6$  & $\bar{c}uc\bar{d}$ $(D\,\bar{D}^*)$ &  \multirow{3}{*}{$4991\pm 815$} & \multirow{3}{*}{$3574\pm {113}$} &$3304\pm {164}$ & $3676\pm {183}$ \\  \cline{1-4}
$Z_c(4200)^\pm$ & $1^+\,( 1^{+-})$	&  $4196^{+35}_{-32}$  & $\bar{c}uc\bar{d}$ &  &  &$3494\pm {136}$  &$3981\pm {195}$\\   \cline{1-4}

$Z_c(4430)^\pm$ & $1^+\,( 1^{+-})$	& $4478^{+15}_{-18}$ & $\bar{c}uc\bar{d}$  $(D_1 D^*\,,\,D_1^\prime D^*)$    & &   &$3611\pm {126}$&$4052\pm {197}$\\  \hline

$Z_b(10610)^\pm$ & $ 1^+\,( 1^{+-})$	& $10607.2\pm2.0$  & $b\bar{d}\bar{b}u$  $(B \bar{B}^*)$ &  \multirow{2}{*}{ $(1.47\pm 0.17) \times 10^4$} & \multirow{2}{*}{$8693\pm {293}$}  & $8481\pm {380}$ & $8918\pm {447}$ \\  \cline{1-4}
$Z_b(10650)^\pm$ & $1^+\,( 1^{+-})$	& $10652.2\pm1.5$  &$b\bar{d}\bar{b}u$ $(B^* \bar{B}^*)$ &  &  & $8550\pm {271}$& $9103\pm {159}$  \\ \hline

$Z_{cs}(4220)^+$ & $1/2\,(1^+)$ & $4216^{+50}_{-40}$  & $u\bar{c}\bar{s}c$ &  $----$  & $3408\pm {234}$ & $3267\pm {136}$ &$3054\pm {182}$ \\ \hline
$R_{c0}(4240)$ & $1^+\,( 0^{--})$ & $4239^{+50}_{-21}$  & $c\bar{u}\bar{c}u$ &  $----$  & $2790\pm {271}$ & $3379\pm {471}$ &$3760\pm {469}$ \\ \hline

\end{tabular}}
	\caption{Our DNN predictions for the mass of some exotic mesons (in units of MeV), 
based on the A1, compared to the experimental \cite{ParticleDataGroup:2022pth}
 and the other NN results in Ref.\cite{Gal:2020dyc}.}\label{tab:exomassA1}
\end{center}
\end{table}

\begin{table}[h!]
\begin{center}
\renewcommand{\arraystretch}{1.4}
\scalebox{0.59}{
\begin{tabular}{|c|c|c|c|c|c|c|c|}\cline{1-8}
Meson&  $I^G\,(J^{PC})$	 & Exp. Mass (MeV) \cite{ParticleDataGroup:2022pth}  & quark content & Ref.\cite{Gal:2020dyc} &base  data &  base data $ +\, h $   &  base data $ +\, h\,G\,C $ \\ \hline
 \hline
$\chi_{c1}(3872)$	 
& $0^+\,(1^{++})$	& $3871.65\pm 0.06$  & $c\bar{u}\bar{c}u$ $(D^0\,\bar{D}^{*0})$ &   $4815\pm 786$  &   $3004\pm540 $   &  $3466\pm673$&  $3345\pm629 $\\ \hline
$\psi(4230)$ & $ 0^-\,( 1^{--})$ & $4222.5\pm 2.4$  & $c\bar{s}\bar{c}s$ $(D_s\,\bar{D}_s)$ &  $(5.4\pm 1.1)\times 10^3$   &   $3613\pm294 $   &  $3313\pm280$&  $3277\pm 379$ \\ \hline
$\psi(4360)$ & $ 0^-\,( 1^{--})$ &  $4374\pm 7$  &  $c\bar{u}\bar{c}u$ $(D_1\,\bar{D}^*)$ & \multirow{2}{*}{$4940\pm 903$} & \multirow{2}{*}{ $3740\pm {609}$} & $3831\pm681$& $3659\pm {606}$ \\ \cline{1-4}
$\psi(4660)$ & $0^-\,( 1^{--})$ & $4630\pm 6$  &  $c\bar{u}\bar{c}u$ $(f_0(980)\,\psi^\prime)$ &  & &  $4186\pm671 $&$3767\pm701 $ \\ \hline
$Z_c(3900)^\pm$ & $ 1^+\,( 1^{+-})$	&$3887.1\pm 2.6$  & $\bar{c}uc\bar{d}$ $(D\,\bar{D}^*)$ &  \multirow{3}{*}{$4991\pm 815$} & \multirow{3}{*}{$2754\pm316$} &$2969\pm108$ & $3455\pm340$ \\  \cline{1-4}
$Z_c(4200)^\pm$ & $1^+\,( 1^{+-})$	&  $4196^{+35}_{-32}$  & $\bar{c}uc\bar{d}$ &  &  &$3012\pm228$  &$3587\pm315$\\   \cline{1-4}

$Z_c(4430)^\pm$ & $1^+\,( 1^{+-})$	& $4478^{+15}_{-18}$ & $\bar{c}uc\bar{d}$  $(D_1 D^*\,,\,D_1^\prime D^*)$    & &   &$3280\pm367$&$3666\pm732$\\  \hline

$Z_b(10610)^\pm$ & $ 1^+\,( 1^{+-})$	& $10607.2\pm2.0$  & $b\bar{d}\bar{b}u$  $(B \bar{B}^*)$ &  \multirow{2}{*}{ $(1.47\pm 0.17)\times10^4$} & \multirow{2}{*}{$9548\pm554$}  & $9634\pm526$ & $9541\pm417$ \\  \cline{1-4}
$Z_b(10650)^\pm$ & $1^+\,( 1^{+-})$	& $10652.2\pm1.5$  &$b\bar{d}\bar{b}u$ $(B^* \bar{B}^*)$ &  &  & $9856\pm500$& $9494\pm484$  \\ \hline

$Z_{cs}(4220)^+$ & $1/2\,(1^+)$ & $4216^{+50}_{-40}$  & $u\bar{c}\bar{s}c$ &  $----$   &   $3610\pm227 $   &  $3975\pm198$&  $3927\pm201 $ \\ \hline
$R_{c0}(4240)$ & $1^+\,( 0^{--})$ & $4239^{+50}_{-21}$  & $c\bar{u}\bar{c}u$ &  $----$  & $2568\pm436$ & $2994\pm554$ &$2816\pm553$ \\ \hline

\end{tabular}
}
	\caption{Our DNN predictions for the mass of some exotic mesons (in units of MeV), 
based on the A2, compared to the experimental \cite{ParticleDataGroup:2022pth}
 and the other NN results in Ref.\cite{Gal:2020dyc}.}\label{tab:exomassA2} 
\end{center}
\end{table}

\begin{table}[h!]
	\begin{center}
		\renewcommand{\arraystretch}{1.4}
		\scalebox{0.68}{
			\begin{tabular}{|c|c|c|c|}\cline{1-4}
				Exotic & A1\% 	 & A2\% &  Ref.\cite{Gal:2020dyc} \% \\ \hline
				\hline  
				base$  $ & 19.87 &22.43 &	 \multirow{3}{*}{23.03}	 	\\  \cline{1-3}
				base$ +\, h $ & 20.42 & 16.76 &\\    \cline{1-3}
				base$ +\, h \,G\,C$ & 17.36 & 16.01 & \\  
				\hline
			\end{tabular}
		}
		\caption{The mean errors of the predicted mass of the exotic mesons based on 
		the A1 and A2 in comparison with Ref.\cite{Gal:2020dyc}.}  \label{tab:exor}
	\end{center}
\end{table} 

\subsection{\textit{Meson width}}
We have trained the well-built DNNs to estimate the mass of various mesons, 
both the ordinary and exotic, based on their quark content and quantum numbers. 
Our DNNs have achieved noticeable results in prediction of the mass.
Now, we are going to estimate the decay width of mesons and tetraquarks 
whose widths have not been experimentally verified or measured.
For this goal, the mesons' mass will be appended to the feature columns in the dataset 
and the same procedure mentioned above, are performed. 
Tables \eqref{tab:widthA1} to \eqref{tab:widthA22}
present the predicted widths of some selected mesons 
according to the A1 and A2 respectively, 
along with the available experimental data for comparison.

As mentioned earlier, the quark content of the $a_0(980)$, 
$f_0(980)$, $D^*_{s0}(2317)^{\pm}$ and $D_{s1}(2460)$ mesons
have turned out to be mysterious.
The DNN is designed to predict the decay width of these challenging mesons too.
We first suppose the $q\bar{q}$ structure and then examine the $q\bar{q}q\bar{q}$ structure for their quark contents.
The predicted widths are shown in tables \eqref{tab:widthA1} and \eqref{tab:widthA2}
for the A1 and A2 respectively.
The enhanced accuracy in estimating the width of the $a_0(980)$ meson, 
in both the A1 and A2, when modeled as an ordinary meson rather than an exotic state, 
lends credence to the conventional $q\bar{q}$ structure over the more complex tetraquark configuration. 
The DNN offers that, the $a_0(980)$ meson is more aptly described within the traditional mesonic framework, highlighting the importance of structure in determining meson characteristics.

The PDG estimates a range from $10$ to  $100$ MeV for the $f_0(980)$ width.
The peak width in $\pi\pi$ is reported about $50$ MeV, but it might be much larger.
The predicted width of $f_0(980)$ by our DNN,
coincides nicely with the estimated range in the PDG, if we regard it the ordinary meson as well. Likewise,
The width estimations for the mesons $D^*_{s0}(2317)$ and $D_{s1}(2460)$, which align with the conventional $q\bar{q}$ structure, fall within the experimentally reported upper bounds. 
In contrast, the predictions based on exotic structures significantly exceed these bounds. 
Just find it interesting that, our DNNs strongly predict these mesons' decay width values 
close to expected ones when the $q\bar{q}$ configuration is considered.
This disparity suggests that the $q\bar{q}$ structure might be more plausible for these mesons, 
reinforcing the traditional mesonic framework over exotic configurations.

The enhanced precision of the second approach in predicting the widths of mesons, such as $K^*_0(700)$ and $K^*_0(1430)$, whose measurements have been experimentally confirmed, is noteworthy. The predictions align most closely with the observed data when the model incorporates the higher state $h$ and the quantum numbers $C$ and $G$. This suggests that a comprehensive model that includes these factors is crucial for accurate theoretical estimations of the meson properties.

The mass and the decay width spectra of some mesons and tetraquarks 
as well as corresponding uncertainty regions estimated through the DNNs
compared to the current review of the PDG \cite{ParticleDataGroup:2022pth} are illustrated in Fig. \eqref{Plt}. 
Left and right panels are based on the A1 and A2, respectively. 
It seems obvious that our results are found to be in fairly good agreement with the experimental data reported by PDG \cite{ParticleDataGroup:2022pth}. 
Especially, when the Clebsch-Gordan coefficients contribute in 
modifying the dataset (A2), 
the DNNs predict more accurate results for the mass and width of mesons. 
As a consequence, the dataset is optimized in an effective way in the A2.

\begin{figure}[h!]
 \begin{center} 
\includegraphics[width=0.49\textwidth]{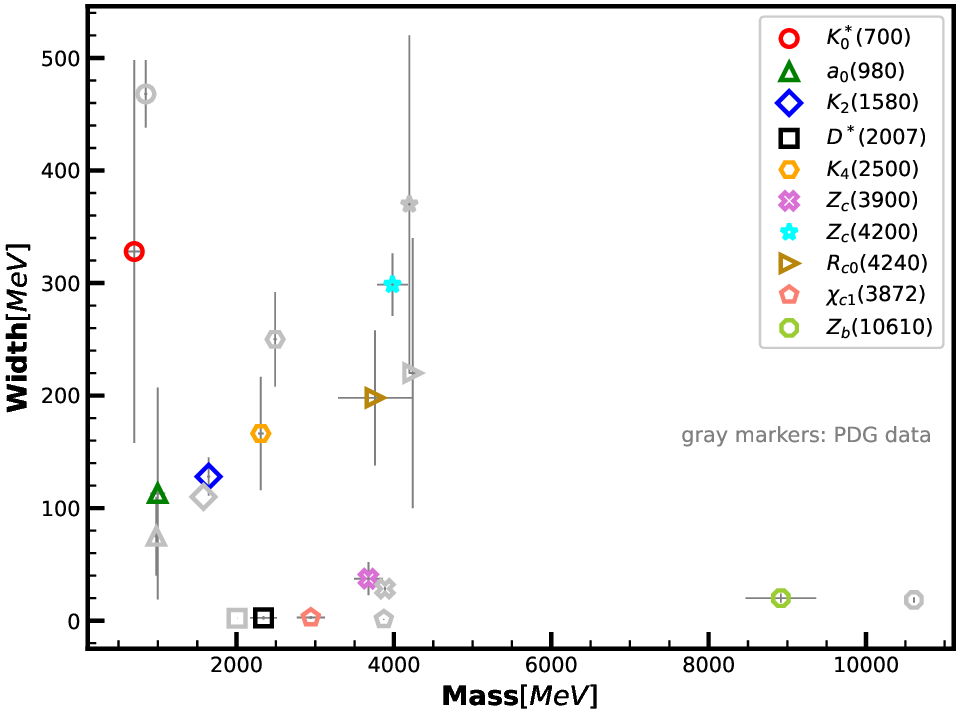}
\includegraphics[width=0.49\textwidth]{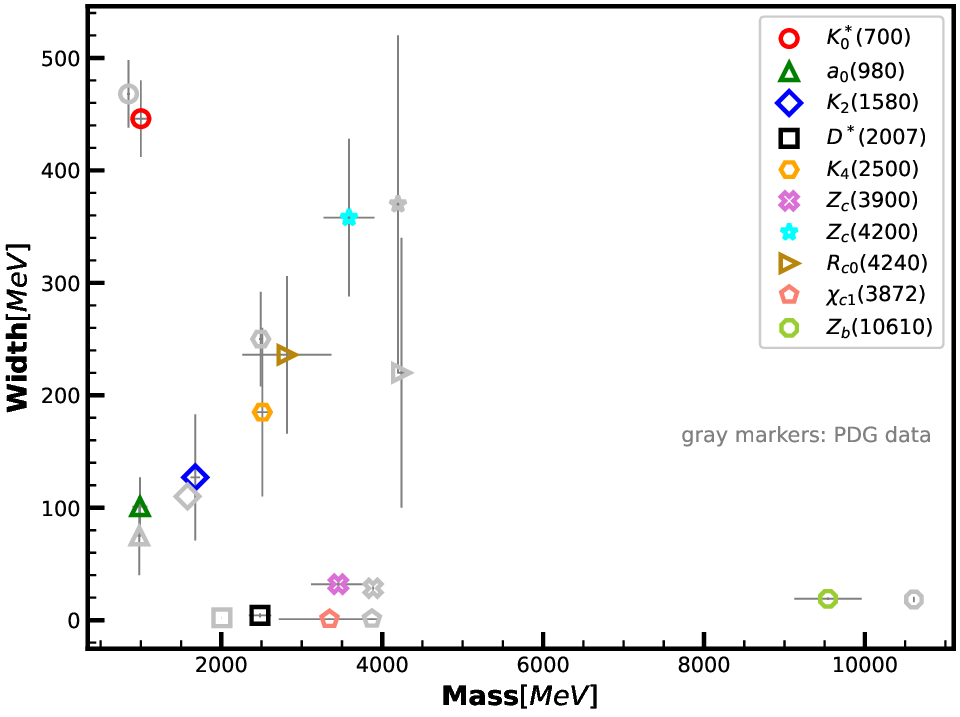}
\end{center}
\caption{The mass and width of some mesons and tetraquarks.
Colorful values with their gray  uncertainties belong to our DNN predictions and 
gray ones (having the same symbols with colorful ones for corresponding particles ) are taken from the PDG  \cite{ParticleDataGroup:2022pth}.
Left panel (Right panel) corresponds to the results, based on the A1 (A2). The uncertainties for our predictions as well as the experimental data are all shown by the gray color to make the figure more readable.}\label{Plt}
\end{figure} 
Besides, tables \eqref{tab:widthA12} and \eqref {tab:widthA22} display
the DNN predictions for the width of some mesons, whose widths have not been experimentally known, 
based on the A1 and A2 respectively. These results may help experimental groups in the course of their search for the corresponding resonances and measure their width.

At the end, tables \eqref{tab:newtetA1} and \eqref{tab:newtetA2} 
demonstrate the predicted mass and width for some new tetraquark states compared to the experiment. 
The congruence of both approaches with the experimentally confirmed masses and widths for the mesons $T^{a++}$ and $T^{a0}$ underscores the reliability of the predictive models within the bounds of uncertainty. Similarly, the consistency of the mass and width predictions for the $Z_V^{+}$ meson with the theoretical calculations presented in Ref. \cite{Agaev:2021jsz}, despite the lack of experimental observation, further validates the robustness of these approaches in the field of particle physics. This alignment between theory and experiment, or calculation in the case of $Z_V^{+}$, is crucial for the ongoing refinement and validation of mesonic models.
The second approach again provides a better estimation for the mass and the width.
The mean error determined,  $4.35 \%$ and $36.36 \%$,  respectively for the mass and  width in the second approach compared to the first approach with $8.60 \%$ and $40.70 \%$ for the errors of mass and width
is depicted in table \eqref{tab:newexr}. 
%Also our estimate of the mass and width of the $Z_V^{++}$ particle is consistent with the results from the QCD Sum Rules technique.

\begin{table}[h!]
\begin{center}
\renewcommand{\arraystretch}{1.4}
\scalebox{0.68}{
\begin{tabular}{|c|c|c|c|c|c|}\cline{1-6}
Meson & $I\,(J^{PC})$	& Width (MeV)&base  data &  base data $ +\, h $   &  base data $ +\, h\,G\,C $ \!\!\! \\ \hline
\hline
$a_0(980)$	 & $1^-\,(0^{++})$	&  $97 \pm 1.9 \pm 5.7$ \cite{CrystalBarrel:2019zqh}  & $158\pm48 $ & 106$\pm34 $& $113\pm28$\\ \hline
$a_0(980)_{exotic}$	 & $1^-\,(0^{++})$	& $97 \pm 1.9 \pm 5.7$ \cite{CrystalBarrel:2019zqh} & $371\pm 90 $ & $320\pm 40 $ & $179\pm 84$\\ \hline

$f_0(980) $	 & $0^+\,(0^{++})$	&$ 10-100 $ \cite{ParticleDataGroup:2022pth}&$114\pm29   $  & $122\pm30 $ & $105\pm34 $\\ \hline
$f_0(980)_{exotic} $	 & $0^+\,(0^{++})$	&$ 10-100 $\cite{ParticleDataGroup:2022pth}&  $253\pm72   $  & $288\pm70 $ & $120\pm 58$\\ \hline
$ f_0(1370)$	 & $0^+\,(0^{++})$	& $200 - 500$ \cite{ParticleDataGroup:2022pth} & $144\pm 24  $  & $132\pm 36 $ & $107\pm 40$\\ \hline
$ D^*(2007)^0$	 & $1/2\,(1^-)$	& $< 2.1$ ($CL = 90 \%$) \cite{Abachi:1988fw} & $1.6\pm 0.8   $  & $4.4\pm 1.8$ & $4.6\pm1.2 $\\ \hline
$ D_{s_0}^{*}(2317)^{\pm}$	 & $0\,(0^+)$	& $ <3.8$ ($CL = 95 \%$) \cite{BaBar:2006eep} & $4.2\pm2   $  & $3.4\pm1.5 $ & $3.1\pm 1.7 $\\ \hline
$D_{s_0}^{*}(2317)^\pm_{exotic} $	 & $0\,(0^+)$	& $ <3.8$ ($CL = 95 \%$) \cite{BaBar:2006eep} & $100\pm46   $  & $23\pm 11 $ & $47\pm 23 $\\ \hline
$ D_{s1}(2460)^{\pm}$	 & $0\,(1^+)$	& $< 3.5$ ($CL = 95 \%$) \cite{BaBar:2006eep} & $3.4\pm2.1   $  & $3.4\pm2.1 $ & $3.4\pm1.6 $\\ \hline
$ D_{s1}(2460)^\pm_{exotic} $	 & $0\,(1^+)$	& $< 3.5$ ($CL = 95 \%$) \cite{BaBar:2006eep} & $85\pm39   $  & $25\pm14 $ & $31\pm 21 $\\ \hline
$\psi_2(3823) $	 & $0\,(2^{--})$	& $ < 2.9$ ($CL = 90 \%$) \cite{BESIII:2022yga} & $5.9\pm 2   $  & $3.7\pm2 $ & $6\pm4 $\\ \hline
$\eta_b(2s) $	 & $0^+\,(0^{-+})$	& $< 24 (CL = 90 \%)$ \cite{Belle:2012fkf}  & $40\pm18   $  & $38\pm21 $ & $54\pm24 $\\ \hline
$ K_0^*(700)$	 & $1/2\,(0^+)$	& $ 468 \pm 30$ \cite{ParticleDataGroup:2022pth} & $177\pm 52  $  & $363\pm 120 $ & $328\pm 140$\\ \hline
$K_0^{*}(1430) $	 & $1/2\,(0^+)$	& $270 \pm 80 $ \cite{ParticleDataGroup:2022pth} & $191\pm 40   $  & $247\pm48 $ & $254\pm 84 $\\ \hline

\end{tabular}
}
	\caption{Our DNN predictions for the width of some mesons (in units of MeV), 
	based on the A1, compared to the experimental results.}\label{tab:widthA1}
\end{center}
\end{table}

\begin{table}[h!]
\begin{center}
\renewcommand{\arraystretch}{1.4}
\scalebox{0.68}{
\begin{tabular}{|c|c|c|c|c|c|}\cline{1-6}
Meson & $I\,(J^{PC})$	& width (MeV) &base  data &  base data $ +\, h $   &  base data $ +\, h\,G\,C $ \!\!\! \\ \hline
\hline
$a_0(980)$	 & $1^-\,(0^{++})$	& $97 \pm 1.9 \pm 5.7$ \cite{CrystalBarrel:2019zqh} & $112\pm 68 $  & $99\pm44 $ & $101\pm26 $\\ \hline
$a_0(980)_{exotic}$	 & $1^-\,(0^{++})$ & $97 \pm 1.9 \pm 5.7$ \cite{CrystalBarrel:2019zqh} & $389\pm115 $  & $552\pm 300$ & $234\pm97 $\\ \hline
$f_0(980) $	 &	 $0^+\,(0^{++})$&$ 10-100 $\cite{ParticleDataGroup:2022pth}&    $115\pm 37$  & $124\pm 42$ & $72\pm44 $\\ \hline
$f_0(980)_{exotic} $	 & $0^+\,(0^{++})$	&$ 10-100 $\cite{ParticleDataGroup:2022pth}&   $303\pm83 $  & $545\pm285 $ & $315\pm185 $\\ \hline
$ f_0(1370)$	 &	 $0^+\,(0^{++})$& $200 - 500$ \cite{ParticleDataGroup:2022pth} & $116\pm16 $  & $119\pm20 $ & $95\pm49 $\\ \hline
$ D^*(2007)^0$	 &  $1/2\,(1^-)$	&  $< 2.1$ ($CL = 90 \%$) \cite{Abachi:1988fw} & $4.8\pm1.4 $  & $4.9\pm1.3 $ & $4.3\pm 1.7$\\ \hline
$ D_{s_0}^{*}(2317)^{\pm}$	 & $0\,(0^+)$	& $ <3.8$ ($CL = 95 \%$) \cite{BaBar:2006eep} & $0.6\pm0.4 $  & $1.75\pm1 $ & $1.9\pm 1.1$\\ \hline
$D_{s_0}^{*}(2317)^{\pm}_{exotic} $	 &$0\,(0^+)$ &  $ <3.8$ ($CL = 95 \%$) \cite{BaBar:2006eep} & $185\pm80 $  & $99\pm52 $ & $140\pm73 $\\ \hline
$ D_{s1}(2460)^{\pm}$	 &  $0\,(1^+)$		&  $< 3.5$ ($CL = 95 \%$) \cite{BaBar:2006eep} & $0.94\pm0.3 $  & $3.7\pm1.9 $ & $6.1\pm4.6 $\\ \hline
$ D_{s1}(2460)^{\pm}_{exotic} $	 &	 $0\,(1^+)$	& $< 3.5$ ($CL = 95 \%$) \cite{BaBar:2006eep} & $83\pm34 $  & $37\pm19 $ & $43\pm20 $\\ \hline
$\psi_2(3823) $	 &  $0\,(2^{--})$	& $ < 2.9$ ($CL = 90 \%$) \cite{BESIII:2022yga} & $2.4\pm1.5 $  & $0.8\pm0.5 $ & $0.6\pm0.1 $\\ \hline
$\eta_b(2s) $	 &$0^+\,(0^{-+})$ &$< 24 (CL = 90 \%)$ \cite{Belle:2012fkf} &$10\pm0.9 $  & $7.8\pm2.7 $ & $11\pm3 $\\ \hline
$ K_0^*(700)$	 &   $1/2\,(0^+)$		&$ 468 \pm 30$ \cite{ParticleDataGroup:2022pth} & $318\pm38 $  & $404\pm44 $ & $446\pm34 $\\ \hline
$K_0^{*}(1430) $	 &  $1/2\,(0^+)$		& $270 \pm 80 $ \cite{ParticleDataGroup:2022pth}  & $239\pm50 $  & $242\pm31 $ & $270\pm40 $\\ \hline

\end{tabular}
}
	\caption{ Our DNN predictions for the width of some mesons (in units of MeV), 
	based on the A2, compared to the experimental results.
}\label{tab:widthA2}
\end{center}
\end{table}

\begin{table}[h!]
\begin{center}
\renewcommand{\arraystretch}{1.4}
\scalebox{0.68}{
\begin{tabular}{|c|c|c|c|c|}\cline{1-5}
Meson & $I\,(J^{PC})$	&base  data &  base data $ +\, h $   &  base data $ +\, h\,G\,C $ \!\!\! \\ \hline
\hline
$B^* $	 & $1/2\,(1^-)$	&   $(2.3 \pm 1.5)\times 10^{-2}  $  & $5.8\pm4.4 $ & $0.84\pm0.56 $\\ \hline
$B^*_{s_0} $	 & $0\,(1^-)$	& $1.39\pm \times 10^{-2} $  & $2.8\pm 1 $ & $0.63\pm 0.2 $\\ \hline
$B_c(2s)^{\pm} $	 & $0\,(0^-)$	& $0.02\pm0.014  $  & $2.3\pm0.8 $ & $2.4\pm 1 $\\ \hline
$ \chi_{b_0}(1p)$	 & $0^+\,(0^{++})$	&   $49\pm 22  $  & $40\pm 19 $ & $45\pm 20 $\\ \hline
$\chi_{b_0}(2p) $	 & $0^+\,(0^{++})$	&   $61\pm 34  $  & $30\pm23 $ & $50\pm 28$\\ \hline
$ \chi_{b_1}(1p)$	 & $0^+\,(1^{++})$	&   $39\pm 19  $  & $19\pm10 $ & $16\pm9 $\\ \hline
$ \chi_{b_1}(2p)$	 & $0^+\,(1^{++})$	&   $47\pm 25   $  & $15\pm 6 $ & $20\pm 10$\\ \hline
$\chi_{b_1}(3p)$	 & $0^+\,(1^{++})$	&   $51\pm 23  $  & $16\pm6 $ & $29\pm19 $\\ \hline
$ \chi_{b_2}(1p)$	 & $0^+\,(2^{++})$	&   $32\pm 19  $  & $21\pm11 $ & $21\pm7 $\\ \hline
$\chi_{b_2}(2p)$	 & $0^+\,(2^{++})$	&   $48\pm20   $  & $16\pm 12$ & $23\pm 11$\\ \hline
$ \chi_{b_2}(3p)$	 & $0^+\,(2^{++})$	&   $42\pm 22  $  & $17\pm 6 $ & $39\pm 19$\\ \hline
$h_b(1p) $	 & $0^-\,(1^{+-})$	&  $44\pm12   $  & $19\pm10 $ & $33\pm16 $\\ \hline
$ h_b(2p)$	 & $0^-\,(1^{+-})$	&  $52\pm 15  $  & $15\pm6 $ & $34\pm12 $\\ \hline
$ K_0,\bar{K}_0$	 & $1/2\,(0^-)$	&   $(2.5 \pm  1.2)\times 10^{-5}   $  & $(2.24\pm1.7)\times 10^{-6} $ & $(3.34\pm 1.8)\times 10^{-6} $\\ \hline
$ \Upsilon_2(1D)$	 & $0^-\,(2^{--})$	&  $41\pm8   $  & $34\pm14 $ & $47\pm 21 $\\ \hline

\end{tabular}
}
	\caption{Our DNN predictions for the width of some mesons (in units of MeV),
	 whose widths have not been experimentally known, 
	based on the A1.}\label{tab:widthA12}
\end{center}
\end{table}

\begin{table}[h!]
\begin{center}
\renewcommand{\arraystretch}{1.4}
\scalebox{0.68}{
\begin{tabular}{|c|c|c|c|c|}\cline{1-5}
Meson & $I\,(J^{PC})$	&base  data &  base data $ +\, h $   &  base data $ +\, h\,G\,C $ \!\!\! \\ \hline
\hline
$ B^{*}$	 & $1/2\,(1^-)$	&    $1.7\pm0.9 $  & $2.4\pm 0.8$ & $2.0\pm0.5 $\\ \hline
$B^*_{s_0} $	 & 	$0\,(1^-)$&   $0.14\pm0.06 $  & $0.058\pm0.025 $ & $0.18\pm0.05 $\\ \hline
$B_c(2s)^{\pm} $	 & $0\,(0^-)$		&    $1.79\pm0.6 $  & $1.8\pm0.5 $ & $2.5\pm0.9 $\\ \hline
$ \chi_{b_0}(1p)$	 &  $0^+\,(0^{++})$		&    $35\pm12 $  & $39\pm10 $ & $23\pm9 $\\ \hline
$\chi_{b_0}(2p) $	 &  $0^+\,(0^{++})$		&  $37\pm11 $  & $37\pm8 $ & $21\pm11 $\\ \hline
$\chi_{b_1}(1p)$	 & $0^+\,(1^{++})$	&    $57\pm20 $  & $4724\pm $ & $57\pm38 $\\ \hline
$ \chi_{b_1}(2p)$	 & $0^+\,(1^{++})$		&   $52\pm17 $  & $48\pm27 $ & $65\pm34 $\\ \hline
$\chi_{b_1}(3p)$	 & $0^+\,(1^{++})$		&   $49\pm19 $  & $42\pm20 $ & $48\pm 17$\\ \hline
$ \chi_{b_2}(1p)$	 & $0^+\,(2^{++})$	&   $81\pm37 $  & $49\pm18 $ & $43\pm17 $\\ \hline
$ \chi_{b_2}(2p)$	 &  $0^+\,(2^{++})$		&    $79\pm34 $  & $43\pm22 $ & $41\pm23 $\\ \hline
$\chi_{b_2}(3p)$	 &  $0^+\,(2^{++})$	&    $75\pm31 $  & $75\pm24 $ & $40\pm30 $\\ \hline
$h_b(1p) $	 &  $0^-\,(1^{+-})$		&   $57\pm19 $  & $47\pm12 $ & $33\pm15 $\\ \hline
$ h_b(2p)$	 &  $0^-\,(1^{+-})$	&    $52\pm17 $  & $43\pm14 $ & $20\pm 10$\\ \hline
$ K_0,\bar{K}_0$	 &  $1/2\,(0^-)$		&   $1.4\pm0.9 $  & $0.7\pm0.3 $ & $0.6\pm0.4 $\\ \hline
$ \Upsilon_2(1D)$	 &  $0^-\,(2^{--})$	&   $92\pm17 $  & $83\pm16 $ & $63\pm30 $\\ \hline

\end{tabular}
}
	\caption{Our DNNs predictions for the width of some mesons (in units of MeV),
	 whose widths have not been experimentally known, 
	based on the A2.
}\label{tab:widthA22}
\end{center}
\end{table}

 \begin{table}[h!]
	\begin{center}
		
		\renewcommand{\arraystretch}{1.4}
		\scalebox{0.68}{
			\begin{tabular}{|c|c|c|c|c|c|c|}\cline{1-7}
				Meson&  $I^G\,(J^{PC})$	 & quark content & Exp. mass (MeV)   & Exp. width (MeV) &mass & width \\ \hline
				\hline
				$Z_V^{++}$ & $1\,(1^{-})$ &$ c\bar{d} u\bar{s}$ & $3515 \pm 125$ \cite{Agaev:2021jsz}* & $156^{+56}_{-30}$ \cite{Agaev:2021jsz}* &$ 2925 \pm	269 $&$ 190\pm	90 $\\ \hline
				$T^{a++}$ & $1\,(0^{+})$ &$ c\bar{d} u\bar{s}\, (D^{ *+} K ^{*+})$&$ 2921\pm17\pm20 $\cite{LHCb:2022sfr} &$ 137\pm32\pm17 $\cite{LHCb:2022sfr}& $ 2788 \pm390 $&$  176\pm	81$\\ \hline
				$T^{a0}$ & $1\,(0^{+})$ &$ c\bar{u} d\bar{s}\, (K^0D^0)$ &$ 2892 \pm 14 \pm15 $ \cite{LHCb:2022sfr}&$ 119\pm26\pm13 $\cite{LHCb:2022sfr}&$ 2526\pm	419 $& $ 182\pm	82 $\\ \hline
				$T^{a+}$ & $1\,(0^{+})$ &$ c\bar{d} d\bar{s}\, (K^0D^+)$& $----$ & $----$& $ 2446	\pm2601 $&$ 186	\pm79 $\\ \hline

			\end{tabular}
		}
	\end{center}\caption{Our DNN predictions for the mass and width of four new tetraquarks based on A1. The asterisk (*) signifies that the data is derived from theoretical calculations in the absence of experimental results. }\label{tab:newtetA1}
\end{table} 

\begin{table}[h!]
	\begin{center}
		
		\renewcommand{\arraystretch}{1.4}
		\scalebox{0.68}{
			\begin{tabular}{|c|c|c|c|c|c|c|}\cline{1-7}
				Meson&  $I^G\,(J^{PC})$	 & quark content & Exp. mass (MeV)  & Exp. width (MeV)& mass & width \\ \hline
				\hline
				$Z_V^{++}$ & $1\,(1^{-})$ &$ c\bar{d} u\bar{s}$& $3515 \pm 125$ \cite{Agaev:2021jsz}* & $156^{+56}_{-30}$ \cite{Agaev:2021jsz}* &$3195\pm 280$&$ 187	\pm 89 $\\ \hline
				$T^{a++}$ & $1\,(0^{+})$ &$ c\bar{d} u\bar{s}\, (D^{ *+} K ^{*+})$&$ 2921\pm17\pm20 $ \cite{LHCb:2022sfr}  &$ 137\pm32\pm17 $\cite{LHCb:2022sfr} &$3158\pm 415$&$ 171 \pm	77 $\\ \hline
				$T^{a0}$ & $1\,(0^{+})$ &$ c\bar{u} d\bar{s}\, (K^0D^0)$   &$ 2892 \pm 14 \pm15 $ \cite{LHCb:2022sfr}&$ 119\pm26\pm13 $\cite{LHCb:2022sfr}& $2875 \pm 536$&$ 176 \pm	78 $\\ \hline
				$T^{a+}$ & $1\,(0^{+})$ & $ c\bar{d} d\bar{s}\, (K^0D^+)$& $----$& $----$&$2704 \pm	338$&$ 182 \pm	77 $\\ \hline

			\end{tabular}
		}
	\end{center}\caption{Our DNN predictions for the mass and width of four new tetraquarks based on the A2. The asterisk (*) signifies that the data is derived from theoretical calculations in the absence of experimental results.}\label{tab:newtetA2}
\end{table}

\begin{table}[h!]
	\begin{center}
		\renewcommand{\arraystretch}{1.4}
		\scalebox{0.68}{
			\begin{tabular}{|c|c|c|}\cline{1-3}
				New exotic & A1 \%	 & A2 \% \\ \hline
				\hline  
				Mass & 8.60 & 4.35 	\\   \hline
				Width  & 40.35 & 36.36\\  
				
				\hline
			\end{tabular}
		}
		\caption{The mean errors of the predicted mass and width of 
		the new tetraquarks based on the A1 and A2.}  \label{tab:newexr}
	\end{center}
\end{table}

\section{Summary and conclusions}\label{SC}
Deep learning algorithms have created a unique opportunity of cooperation between the ML and the particle physics communities. 
The application of the DNN in the hadronic physics is progressing well. 

In this study, the DNNs have been trained to predict the mass and width of some ordinary and exotic mesons based on their quark contents and corresponding quantum numbers. 
The quark content of some famous mesons are still controversial, 
whether they are ordinary mesons or exotic states. 
We tried to estimate their mass through the DNN, considering the $ q\bar{q} $  structure and then the $q\bar{q}q\bar{q}$  one for their quark contents separately. 
We found that when the ordinary structure of $ q\bar{q} $ is  employed,
 our DNN predictions for these challenging mesons’ mass are more consistent with the experimental values. 
Furthermore, we evaluated the power of our designed DNNs 
in computing the mass of some light and exotic mesons. 

For the first time, the DNNs are implemented to predict the width of some ordinary and 
exotic mesons whose  widths have not been permanently confirmed by the PDG \cite{ParticleDataGroup:2022pth}. 
We precisely extracted the dataset containing the light and heavy mesons  as well as some tetraquarks. 
The mesons have been categorized based on their quark content  structures and quantum numbers like the  angular momentum, charge conjugation, isospin and parity.   
It is important to stress that, we have presented and tested two approaches for organizing the dataset. 
In general, the obtained results are in reasonable agreement with the experimental values verified by the PDG. 
Especially, when the Clebsch-Gordan coefficients contribute in modifying the dataset in the second approach, 
the DNNs predict more accurate results for the mass and decay width spectra of the mesons and tetraquarks. 
As a consequence, the dataset has been optimized in an effective way in the second approach.

The DNN performance seems to be more stable compared to the common fitting methods, particularly, for low-statistic data. This presented work has the potential to be upgraded in order to accurately 
predict the mass spectra and decay widths of other tetraquarks as well as hybrid mesons.
A next important step will be to explore prominent features of the baryons, pentaquarks
and possible molecular dibaryons through deep learning techniques.
Recently, many new exotic hadrons have been discovered by the experiments.
Comprehensive studies have been made to specify their properties
and structures and it is still a hot  topic and an active area of research.
It would be a worthwhile idea to probe and predict the quantum numbers 
of exotic hadronic states via constructing appropriate DNNs.

Ultimately, calculation and prediction of basic properties of particles including the mass, 
width and the quantum numbers, 
using the statistical learning algorithms, create new path forward toward robust and reliable methods for discovering the fundamental structure of nature.

\section*{ACKNOWLEDGMENTS}

 M. M and S. R   are  grateful to the organizers of 
the MITP Summer School on ``Machine Learning in Particle Theory" for their support. K. A.  thanks  Iran national science foundation (INSF)
for the partial financial support provided under the elites Grant No. 4025036.


\newpage
\begin{thebibliography}{0}
	
	\bibitem{Gell-Mann:1964ewy}
M.~Gell-Mann,
``A Schematic Model of Baryons and Mesons,''
\href{https://doi.org/10.1016/S0031-9163(64)92001-3}{Phys. Lett. \textbf{8}, 214-215 (1964)}.
%doi:10.1016/S0031-9163(64)92001-3

\bibitem{Gell-Mann:1962yej}
M.~Gell-Mann,
``Symmetries of baryons and mesons,''
\href{https://doi.org/10.1103/PhysRev.125.1067}{Phys. Rev. \textbf{125}, 1067-1084 (1962)}.
%doi:10.1103/PhysRev.125.1067

\bibitem{Chodos:1974je}
A.~Chodos, R.~L.~Jaffe, K.~Johnson, C.~B.~Thorn and V.~F.~Weisskopf,
``A New Extended Model of Hadrons,''
\href{https://doi.org/10.1103/PhysRevD.9.3471}{Phys. Rev. D \textbf{9}, 3471-3495 (1974)}.
%doi:10.1103/PhysRevD.9.3471
	
	\bibitem{Griffiths:1983ah}
L.~A.~Griffiths, C.~Michael and P.~E.~L.~Rakow,
``Mesons With Excited Glue,''
\href{https://doi.org/10.1016/0370-2693(83)90680-9}{Phys. Lett. B \textbf{129}, 351-356 (1983)}.
%doi:10.1016/0370-2693(83)90680-9

\bibitem{Chen:2022asf}
H.~X.~Chen, W.~Chen, X.~Liu, Y.~R.~Liu and S.~L.~Zhu,
``An updated review of the new hadron states,''
\href{https://doi.org/10.1088/1361-6633/aca3b6}{Rept. Prog. Phys. \textbf{86}, no.2, 026201 (2023)},
%doi:10.1088/1361-6633/aca3b6
\href{https://arxiv.org/abs/2204.02649}{[arXiv:2204.02649 [hep-ph]]}.
	
	\bibitem{Schumacher:2018evl}
M.~Schumacher,
``Mass and structure of the nucleon: Gluon trace anomaly versus spontaneous symmetry breaking,''
\href{https://arxiv.org/abs/1807.10798}{[arXiv:1807.10798 [hep-ph]]}.

\bibitem{Schumacher:2015wla}
M.~Schumacher,
``Mass generation via the Higgs boson and the quark condensate of the QCD vacuum,''
\href{https://doi.org/10.1007/s12043-016-1256-0}{Pramana \textbf{87}, no.3, 44 (2016)},
%doi:10.1007/s12043-016-1256-0
\href{https://arxiv.org/abs/1506.00410}{[arXiv:1506.00410 [hep-ph]]}.

%\cite{Jaffe:1976ig}
\bibitem{Jaffe:1976ig}
R.~L.~Jaffe,
``Multi-Quark Hadrons. 1. The Phenomenology of (2 Quark 2 anti-Quark) Mesons,''
\href{https://doi.org/10.1103/PhysRevD.15.267}{Phys. Rev. D \textbf{15}, 267 (1977)}.
%doi:10.1103/PhysRevD.15.267

%\cite{Jaffe:1976ih}
\bibitem{Jaffe:1976ih}
R.~L.~Jaffe,
``Multi-Quark Hadrons. 2. Methods,''
\href{https://doi.org/10.1103/PhysRevD.15.281}{Phys. Rev. D \textbf{15}, 281 (1977)}.
%doi:10.1103/PhysRevD.15.281

\bibitem{Dong:2019ofp}
X.~K.~Dong, Y.~H.~Lin and B.~S.~Zou,
``Prediction of an exotic state around 4240 MeV with $J^{PC}=1^{-+}$ as C-parity partner of $ Y(4260) $ in molecular picture,''
\href{https://doi.org/10.1103/PhysRevD.101.076003}{Phys. Rev. D \textbf{101}, no.7, 076003 (2020)},
%doi:10.1103/PhysRevD.101.076003
\href{https://arxiv.org/abs/1910.14455}{[arXiv:1910.14455 [hep-ph]]}.

\bibitem{Voloshin:1976ap}
M.~B.~Voloshin and L.~B.~Okun,
``Hadron Molecules and Charmonium Atom,''
JETP Lett. \textbf{23}, 333-336 (1976).

\bibitem{Esposito:2016noz}
A.~Esposito, A.~Pilloni and A.~D.~Polosa,
``Multiquark Resonances,''
\href{https://doi.org/10.1016/j.physrep.2016.11.002}{Phys. Rept. \textbf{668}, 1-97 (2017)},
%doi:10.1016/j.physrep.2016.11.002
\href{https://arxiv.org/abs/1611.07920}{[arXiv:1611.07920 [hep-ph]]}.

\bibitem{Shifman:1978bx}
M.~A.~Shifman, A.~I.~Vainshtein and V.~I.~Zakharov,
``QCD and Resonance Physics. Theoretical Foundations,''
\href{https://doi.org/10.1016/0550-3213(79)90022-1}{Nucl. Phys. B \textbf{147}, 385-447 (1979)}.
%doi:10.1016/0550-3213(79)90022-1

%\cite{Agaev:2016dev}
\bibitem{Agaev:2016dev}
S.~S.~Agaev, K.~Azizi and H.~Sundu,
``Strong $Z_c^{+}(3900)\rightarrow J/\psi \pi^{+}; \eta_{c} \rho^{+}$ decays in QCD,''
\href{https://doi.org/10.1103/PhysRevD.93.074002}{Phys. Rev. D \textbf{93}, no.7, 074002 (2016)},
%doi:10.1103/PhysRevD.93.074002
\href{https://arxiv.org/abs/1601.03847}{[arXiv:1601.03847 [hep-ph]]}.

%\cite{Agaev:2016srl}
\bibitem{Agaev:2016srl}
S.~S.~Agaev, K.~Azizi and H.~Sundu,
``Application of the QCD light cone sum rule to tetraquarks: the strong vertices $X_bX_b\rho$ and $X_cX_c\rho$,''
\href{https://doi.org/10.1103/PhysRevD.93.114036}{Phys. Rev. D \textbf{93}, no.11, 114036 (2016)},
%doi:10.1103/PhysRevD.93.114036
\href{https://arxiv.org/abs/1605.02496}{[arXiv:1605.02496 [hep-ph]]}.

%\cite{Agaev:2020zad}
\bibitem{Agaev:2020zad}
S.~Agaev, K.~Azizi and H.~Sundu,
``Four-quark exotic mesons,''
\href{https://doi.org/10.3906/fiz-2003-15}{Turk. J. Phys. \textbf{44}, no.2, 95-173 (2020)},
%doi:10.3906/fiz-2003-15
\href{https://arxiv.org/abs/2004.12079}{[arXiv:2004.12079 [hep-ph]]}.

%\cite{Azizi:2019xla}
\bibitem{Azizi:2019xla}
K.~Azizi, S.~S.~Agaev and H.~Sundu,
``The Scalar Hexaquark $uuddss$: a Candidate to Dark Matter?,''
\href{https://doi.org/10.1088/1361-6471/ab9a0e}{J. Phys. G \textbf{47}, no.9, 095001 (2020)},
%doi:10.1088/1361-6471/ab9a0e
\href{https://arxiv.org/abs/1904.09913}{[arXiv:1904.09913 [hep-ph]]}.

\bibitem{Cowan:2016kjn}
G.~A.~Cowan,
``Exotic hadron spectroscopy at the LHCb experiment,''
\href{https://arxiv.org/abs/1610.04906}{[arXiv:1610.04906 [hep-ex]]}.

\bibitem{LHCb:2015yax}
R.~Aaij \textit{et al.} [LHCb],
``Observation of $J/\psi p$ Resonances Consistent with Pentaquark States in $\Lambda_b^0 \to J/\psi K^- p$ Decays,''
\href{https://doi.org/10.1103/PhysRevLett.115.072001}{Phys. Rev. Lett. \textbf{115}, 072001 (2015)},
%doi:10.1103/PhysRevLett.115.072001
\href{https://arxiv.org/abs/1507.03414}{[arXiv:1507.03414 [hep-ex]]}.


\bibitem{LHCb:2016lve}
R.~Aaij \textit{et al}. [LHCb],
``Evidence for exotic hadron contributions to $\Lambda_b^0 \to J/\psi p \pi^-$ decays,''
\href{https://doi.org/10.1103/PhysRevLett.117.082003}{Phys. Rev. Lett. \textbf{117}, no.8, 082003 (2016)},
%doi:10.1103/PhysRevLett.117.082003
\href{https://arxiv.org/abs/1606.06999 }{[arXiv:1606.06999 [hep-ex]]}.

\bibitem{Guo:2015umn}
F.~K.~Guo, U.~G.~Mei\ss{}ner, W.~Wang and Z.~Yang,
``How to reveal the exotic nature of the P$_c$(4450),''
\href{https://doi.org/10.1103/PhysRevD.92.071502}{Phys. Rev. D \textbf{92}, no.7, 071502 (2015)},
%doi:10.1103/PhysRevD.92.071502
\href{https://arxiv.org/abs/1507.04950}{[arXiv:1507.04950 [hep-ph]]}.

\bibitem{LHCb:2022sfr}
R.~Aaij \textit{et al}. [LHCb],
``First Observation of a Doubly Charged Tetraquark and Its Neutral Partner,''
\href{https://doi.org/10.1103/PhysRevLett.131.041902}{Phys. Rev. Lett. \textbf{131}, no.4, 041902 (2023)},
%doi:10.1103/PhysRevLett.131.041902
\href{https://arxiv.org/abs/2212.02716}{[arXiv:2212.02716 [hep-ex]]}.

\bibitem{LHCb:2016axx}
R.~Aaij \textit{et al}. [LHCb],
``Observation of $J/\psi\phi$ structures consistent with exotic states from amplitude analysis of $B^+\to J/\psi \phi K^+$ decays,''
\href{https://doi.org/10.1103/PhysRevLett.118.022003}{Phys. Rev. Lett. \textbf{118}, no.2, 022003 (2017)},
%doi:10.1103/PhysRevLett.118.022003
\href{https://arxiv.org/abs/1606.07895}{[arXiv:1606.07895 [hep-ex]]}.

%\cite{CDF:2003cab}
\bibitem{CDF:2003cab}
D.~Acosta \textit{et al.} [CDF],
``Observation of the narrow state $X(3872) \to J/\psi \pi^+ \pi^-$ in $\bar{p}p$ collisions at $\sqrt{s} = 1.96$ TeV,''
\href{https://doi.org/10.1103/PhysRevLett.93.072001}{Phys. Rev. Lett. \textbf{93}, 072001 (2004)},
%doi:10.1103/PhysRevLett.93.072001
\href{https://arxiv.org/abs/hep-ex/0312021}{[arXiv:hep-ex/0312021 [hep-ex]]}.

%\cite{Belle:2003nnu}
\bibitem{Belle:2003nnu}
S.~K.~Choi \textit{et al.} [Belle],
``Observation of a narrow charmonium-like state in exclusive $B^\pm \to K^\pm \pi^+ \pi^- J/\psi$ decays,''
\href{https://doi.org/10.1103/PhysRevLett.91.262001}{Phys. Rev. Lett. \textbf{91}, 262001 (2003)},
%doi:10.1103/PhysRevLett.91.262001
\href{https://arxiv.org/abs/hep-ex/0309032}{[arXiv:hep-ex/0309032 [hep-ex]]}.

%\cite{CMS:2023owd}
\bibitem{CMS:2023owd}
A.~Hayrapetyan \textit{et al.} [CMS],
``Observation of new structure in the J/$\psi$J/$\psi$ mass spectrum in proton-proton collisions at $\sqrt{s}$ = 13 TeV,''
\href{https://arxiv.org/abs/2306.07164}{[arXiv:2306.07164 [hep-ex]]}.

%\cite{LHCb:2021vvq}
\bibitem{LHCb:2021vvq}
R.~Aaij \textit{et al.} [LHCb],
``Observation of an exotic narrow doubly charmed tetraquark,''
\href{https://doi.org/10.1038/s41567-022-01614-y}{Nature Phys. \textbf{18}, no.7, 751-754 (2022)},
%doi:10.1038/s41567-022-01614-y
\href{https://arxiv.org/abs/2109.01038}{[arXiv:2109.01038 [hep-ex]]}.

%\cite{LHCb:2021auc}
\bibitem{LHCb:2021auc}
R.~Aaij \textit{et al.} [LHCb],
``Study of the doubly charmed tetraquark $T_{cc}^{+}$,''
\href{https://doi.org/10.1038/s41467-022-30206-w}{Nature Commun. \textbf{13}, no.1, 3351 (2022)},
%doi:10.1038/s41467-022-30206-w
\href{https://arxiv.org/abs/2109.01056}{[arXiv:2109.01056 [hep-ex]]}.




\bibitem{Choe:1996uc}
S.~Choe,
``Multi - quark states and QCD sum rules,''
Soryushiron Kenkyu \textbf{95}, D87 (1997)
\href{https://arxiv.org/abs/9705419}{[arXiv:hep-ph/9705419 [hep-ph]]}.

\bibitem{Agaev:2023tzi}
S.~S.~Agaev, K.~Azizi, B.~Barsbay and H.~Sundu,
``Scalar exotic mesons $bb\overline{c}\overline{c}$,''
\href{https://arxiv.org/abs/2311.10534}{[arXiv:2311.10534 [hep-ph]]}.

\bibitem{Agaev:2023wua}
S.~S.~Agaev, K.~Azizi, B.~Barsbay and H.~Sundu,
``Exploring fully heavy scalar tetraquarks $QQ\bar{Q}\bar{Q}$,''
\href{https://doi.org/10.1016/j.physletb.2023.138089}{Phys. Lett. B \textbf{844}, 138089 (2023)},
%doi:10.1016/j.physletb.2023.138089
\href{https://arxiv.org/abs/2304.03244}{[arXiv:2304.03244 [hep-ph]]}.

\bibitem{Agaev:2023ruu}
S.~S.~Agaev, K.~Azizi, B.~Barsbay and H.~Sundu,
``Hadronic molecules $\eta _c \eta _c$ and $\chi _{c0}\chi _{c0}$,''
\href{https://doi.org/10.1140/epjp/s13360-023-04562-5}{Eur. Phys. J. Plus \textbf{138}, no.10, 935 (2023)},
%doi:10.1140/epjp/s13360-023-04562-5
\href{https://arxiv.org/abs/2305.03696}{[arXiv:2305.03696 [hep-ph]]}.

\bibitem{HadronSpectrum:2012gic}
L.~Liu \textit{et al}. [Hadron Spectrum],
``Excited and exotic charmonium spectroscopy from lattice QCD,''
\href{https://doi.org/10.1007/JHEP07(2012)126}{JHEP \textbf{07}, 126 (2012)},
%doi:10.1007/JHEP07(2012)126
\href{https://arxiv.org/abs/1204.5425}{[arXiv:1204.5425 [hep-ph]]}.


\bibitem{Dudek:2009qf}
J.~J.~Dudek, R.~G.~Edwards, M.~J.~Peardon, D.~G.~Richards and C.~E.~Thomas,
``Highly excited and exotic meson spectrum from dynamical lattice QCD,''
\href{https://doi.org/10.1103/PhysRevLett.103.262001}{Phys. Rev. Lett. \textbf{103}, 262001 (2009)},
%doi:10.1103/PhysRevLett.103.262001
\href{https://arxiv.org/abs/0909.0200}{[arXiv:0909.0200 [hep-ph]]}.

\bibitem{Bouhova-Thacker:2022vnt}
E.~Bouhova-Thacker [ATLAS],
``ATLAS results on exotic hadronic resonances,''
\href{https://doi.org/10.22323/1.414.0806}{PoS \textbf{ICHEP2022}, 806 (2022)}.
%doi:10.22323/1.414.0806


\bibitem{ParticleDataGroup:2022pth}
R.~L.~Workman \textit{et al}. [Particle Data Group],
``Review of Particle Physics,''
\href{https://doi.org/10.1093/ptep/ptac097}{PTEP \textbf{2022}, 083C01 (2022)}.
%doi:10.1093/ptep/ptac097



\bibitem{VanHove:1974wa}
L.~Van Hove and S.~Pokorski,
%``High-Energy Hadron-Hadron Collisions and Internal Hadron Structure,''
\href{https://doi.org/10.1016/0550-3213(75)90443-5}{Nucl. Phys. B \textbf{86}, 243-252 (1975)}.
%doi:10.1016/0550-3213(75)90443-5


\bibitem{Briceno:2015rlt}
R.~A.~Briceno, T.~D.~Cohen, S.~Coito, J.~J.~Dudek, E.~Eichten, C.~S.~Fischer, M.~Fritsch, W.~Gradl, A.~Jackura and M.~Kornicer, \textit{et al}.
``Issues and Opportunities in Exotic Hadrons,''
\href{https://doi.org/10.1088/1674-1137/40/4/042001}{Chin. Phys. C \textbf{40}, no.4, 042001 (2016)},
%doi:10.1088/1674-1137/40/4/042001
\href{https://arxiv.org/abs/1511.06779}{[arXiv:1511.06779 [hep-ph]]}.

\bibitem{Calafiura:2022ges}
P.~Calafiura, D.~Rousseau and K.~Terao,
``Artificial Intelligence for High Energy Physics,''
\href{https://doi.org/10.1142/12200}{World Scientific, 2022,
ISBN 978-981-12-3402-6, 978-981-12-3404-0}.
%doi:10.1142/12200

\bibitem{Schwartz:2021ftp}
M.~D.~Schwartz,
``Modern Machine Learning and Particle Physics,''
\href{https://doi.org/10.1162/99608f92.beeb1183}{}
%doi:10.1162/99608f92.beeb1183
\href{https://arxiv.org/abs/2103.12226}{[arXiv:2103.12226 [hep-ph]]}.

\bibitem{Guest:2018yhq}
D.~Guest, K.~Cranmer and D.~Whiteson,
``Deep Learning and its Application to LHC Physics,''
\href{https://doi.org/10.1146/annurev-nucl-101917-021019}{Ann. Rev. Nucl. Part. Sci. \textbf{68}, 161-181 (2018)},
%doi:10.1146/annurev-nucl-101917-021019
\href{https://arxiv.org/abs/1806.11484}{[arXiv:1806.11484 [hep-ex]]}.

\bibitem{Albertsson:2018maf}
K.~Albertsson, P.~Altoe, D.~Anderson, J.~Anderson, M.~Andrews, J.~P.~Araque Espinosa, A.~Aurisano, L.~Basara, A.~Bevan and W.~Bhimji, \textit{et al}.
``Machine Learning in High Energy Physics Community White Paper,''
\href{https://doi.org/10.1088/1742-6596/1085/2/022008}{J. Phys. Conf. Ser. \textbf{1085}, no.2, 022008 (2018)},
%doi:10.1088/1742-6596/1085/2/022008
\href{https://arxiv.org/abs/1807.02876}{[arXiv:1807.02876 [physics.comp-ph]]}.


\bibitem{Duarte:2022job}
J.~Duarte, H.~Li, A.~Roy, R.~Zhu, E.~A.~Huerta, D.~Diaz, P.~Harris, R.~Kansal, D.~S.~Katz and I.~H.~Kavoori, \textit{et al}.
``FAIR AI models in high energy physics,''
\href{https://doi.org/10.1088/2632-2153/ad12e3}{Mach. Learn. Sci. Tech. \textbf{4}, no.4, 045062 (2023)},
%doi:10.1088/2632-2153/ad12e3
\href{https://arxiv.org/abs/2212.05081}{[arXiv:2212.05081 [hep-ex]]}.


\bibitem{CMS:2012qbp}
S.~Chatrchyan \textit{et al}. [CMS],
``Observation of a New Boson at a Mass of 125 GeV with the CMS Experiment at the LHC,''
\href{https://doi.org/10.1016/j.physletb.2012.08.021}{Phys. Lett. B \textbf{716}, 30-61 (2012)},
%doi:10.1016/j.physletb.2012.08.021
\href{https://arxiv.org/abs/1207.7235}{[arXiv:1207.7235 [hep-ex]]}.


\bibitem{CMS:2018nsn}
A.~M.~Sirunyan \textit{et al}. [CMS],
``Observation of Higgs boson decay to bottom quarks,''
\href{https://doi.org/10.1103/PhysRevLett.121.121801}{Phys. Rev. Lett. \textbf{121}, no.12, 121801 (2018)},
%doi:10.1103/PhysRevLett.121.121801
\href{https://arxiv.org/abs/1808.08242}{[arXiv:1808.08242 [hep-ex]]}.

\bibitem{ATLAS:2012yve}
G.~Aad \textit{et al}. [ATLAS],
``Observation of a new particle in the search for the Standard Model Higgs boson with the ATLAS detector at the LHC,''
\href{https://doi.org/10.1016/j.physletb.2012.08.020}{Phys. Lett. B \textbf{716}, 1-29 (2012)},
%doi:10.1016/j.physletb.2012.08.020
\href{https://arxiv.org/abs/1207.7214}{[arXiv:1207.7214 [hep-ex]]}.


\bibitem{Larkoski:2017jix}
A.~J.~Larkoski, I.~Moult and B.~Nachman,
``Jet Substructure at the Large Hadron Collider: A Review of Recent Advances in Theory and Machine Learning,''
\href{https://doi.org/10.1016/j.physrep.2019.11.001}{Phys. Rept. \textbf{841}, 1-63 (2020)},
%doi:10.1016/j.physrep.2019.11.001
\href{https://arxiv.org/abs/1709.04464}{[arXiv:1709.04464 [hep-ph]]}.



\bibitem{Radovic:2018dip}
A.~Radovic, M.~Williams, D.~Rousseau, M.~Kagan, D.~Bonacorsi, A.~Himmel, A.~Aurisano, K.~Terao and T.~Wongjirad,
``Machine learning at the energy and intensity frontiers of particle physics,''
\href{https://doi.org/10.1038/s41586-018-0361-2}{Nature \textbf{560}, no.7716, 41-48 (2018)}.
%doi:10.1038/s41586-018-0361-2

\bibitem{Carleo:2019ptp}
G.~Carleo, I.~Cirac, K.~Cranmer, L.~Daudet, M.~Schuld, N.~Tishby, L.~Vogt-Maranto and L.~Zdeborov\'a,
``Machine learning and the physical sciences,''
\href{https://doi.org/10.1103/RevModPhys.91.045002}{Rev. Mod. Phys. \textbf{91}, no.4, 045002 (2019)},
%doi:10.1103/RevModPhys.91.045002
\href{https://arxiv.org/abs/1903.10563}{[arXiv:1903.10563 [physics.comp-ph]]}.


\bibitem{Bourilkov:2019yoi}
D.~Bourilkov,
``Machine and Deep Learning Applications in Particle Physics,''
\href{https://doi.org/10.1142/S0217751X19300199}{Int. J. Mod. Phys. A \textbf{34}, no.35, 1930019 (2020)},
%doi:10.1142/S0217751X19300199
\href{https://arxiv.org/abs/1912.08245}{[arXiv:1912.08245 [physics.data-an]]}.


\bibitem{Feickert:2021ajf}
M.~Feickert and B.~Nachman,
``A Living Review of Machine Learning for Particle Physics,''
\href{https://arxiv.org/abs/2102.02770}{[arXiv:2102.02770 [hep-ph]]}.

\bibitem{Karagiorgi:2021ngt}
G.~Karagiorgi, G.~Kasieczka, S.~Kravitz, B.~Nachman and D.~Shih,
``Machine Learning in the Search for New Fundamental Physics,''
\href{https://arxiv.org/abs/2112.03769}{[arXiv:2112.03769 [hep-ph]]}.


\bibitem{Shanahan:2022ifi}
P.~Shanahan, K.~Terao, D.~Whiteson, G.~Aarts, A.~Adelmann, N.~Akchurin, A.~Alexandru, O.~Amram, A.~Andreassen and A.~Apresyan, \textit{et al}.
``Snowmass 2021 Computational Frontier CompF03 Topical Group Report: Machine Learning,''
\href{https://arxiv.org/abs/2209.07559 }{[arXiv:2209.07559 [physics.comp-ph]]}.


\bibitem{Plehn:2022ftl}
T.~Plehn, A.~Butter, B.~Dillon and C.~Krause,
``Modern Machine Learning for LHC Physicists,''
\href{https://arxiv.org/abs/2211.01421}{[arXiv:2211.01421 [hep-ph]]}.


\bibitem{Butter:2022rso}
A.~Butter, T.~Plehn, S.~Schumann, S.~Badger, S.~Caron, K.~Cranmer, F.~A.~Di Bello, E.~Dreyer, S.~Forte and S.~Ganguly, \textit{et al}.
``Machine learning and LHC event generation,''
\href{https://doi.org/10.21468/SciPostPhys.14.4.079}{SciPost Phys. \textbf{14}, no.4, 079 (2023)},
%doi:10.21468/SciPostPhys.14.4.079
\href{https://arxiv.org/abs/2203.07460}{[arXiv:2203.07460 [hep-ph]]}.

\bibitem{CMS:2017wtu}
A.~M.~Sirunyan \textit{et al}. [CMS],
``Identification of heavy-flavour jets with the CMS detector in pp collisions at 13 TeV,''
\href{https://doi.org/10.1088/1748-0221/13/05/P05011}{JINST \textbf{13}, no.05, P05011 (2018)},
%doi:10.1088/1748-0221/13/05/P05011
\href{https://arxiv.org/abs/1712.07158}{[arXiv:1712.07158 [physics.ins-det]]}.

\bibitem{ATLAS:2015yey}
G.~Aad \textit{et al}. [ATLAS and CMS],
``Combined Measurement of the Higgs Boson Mass in $pp$ Collisions at $\sqrt{s}=7$ and 8 TeV with the ATLAS and CMS Experiments,''
\href{https://doi.org/10.1103/PhysRevLett.114.191803}{Phys. Rev. Lett. \textbf{114}, 191803 (2015)},
%doi:10.1103/PhysRevLett.114.191803
\href{https://arxiv.org/abs/1503.07589}{[arXiv:1503.07589 [hep-ex]]}.

\bibitem{Pata:2022wam}
J.~Pata \textit{et al}. [CMS],
``Machine Learning for Particle Flow Reconstruction at CMS,''
\href{https://doi.org/10.1088/1742-6596/2438/1/012100}{J. Phys. Conf. Ser. \textbf{2438}, no.1, 012100 (2023)},
%doi:10.1088/1742-6596/2438/1/012100
\href{https://arxiv.org/abs/2203.00330 }{[arXiv:2203.00330 [physics.data-an]]}.

\bibitem{Stoye:2018qgr}
M.~Stoye [CMS],
``Deep learning in jet reconstruction at CMS,''
\href{https://doi.org/10.1088/1742-6596/1085/4/042029}{J. Phys. Conf. Ser. \textbf{1085}, no.4, 042029 (2018)}.
%doi:10.1088/1742-6596/1085/4/042029

\bibitem{Wachirapusitan:2023ttc}
V.~Wachirapusitan [CMS],
``Machine Learning applications for Data Quality Monitoring and Data Certification within CMS,''
\href{https://doi.org/10.1088/1742-6596/2438/1/012098}{J. Phys. Conf. Ser. \textbf{2438}, no.1, 012098 (2023)}.
%doi:10.1088/1742-6596/2438/1/012098


\bibitem{Schramm:2018knx}
S.~Schramm [ATLAS and LHCb],
``Machine learning at CERN: ATLAS, LHCb, and more,''
\href{https://doi.org/10.22323/1.340.0158}{PoS \textbf{ICHEP2018}, 158 (2019)}.
%doi:10.22323/1.340.0158


\bibitem{DeCian:2017ytk}
M.~De Cian, S.~Farry, P.~Seyfert and S.~Stahl,
``Fast neural-net based fake track rejection in the LHCb reconstruction,''
LHCb-PUB-2017-011.



\bibitem{ATLAS:2017gpy}
 [ATLAS],
``Identification of Jets Containing $b$-Hadrons with Recurrent Neural Networks at the ATLAS Experiment,''
ATL-PHYS-PUB-2017-003.


\bibitem{Keicher:2023mer}
P.~Keicher,
``Machine Learning in Top Physics in the ATLAS and CMS Collaborations,''
\href{https://arxiv.org/abs/2301.09534}{[arXiv:2301.09534 [hep-ex]]}.


\bibitem{ATLAS:2023ixc}ATLAS:2023ixc,ATLAS:2015jge,Graczykowski:2022zae
G.~Aad \textit{et al}. [ATLAS],
``Search for new phenomena in two-body invariant mass distributions using unsupervised machine learning for anomaly detection at $\sqrt{s} = 13$ TeV with the ATLAS detector,''
\href{https://arxiv.org/abs/2307.01612}{[arXiv:2307.01612 [hep-ex]]}.


\bibitem{ATLAS:2015jge}
 [ATLAS],
``Vertex Reconstruction Performance of the ATLAS Detector at $\sqrt{s} = 13 $ TeV,''
ATL-PHYS-PUB-2015-026.

\bibitem{Graczykowski:2022zae}
\L{}.~K.~Graczykowski \textit{et al}. [ALICE],
``Using machine learning for particle identification in ALICE,''
\href{https://doi.org/10.1088/1748-0221/17/07/C07016}{JINST \textbf{17}, no.07, C07016 (2022)},
%doi:10.1088/1748-0221/17/07/C07016
\href{https://arxiv.org/abs/2204.06900}{[arXiv:2204.06900 [nucl-ex]]}.

\bibitem{Bas}
Bas~van ~Stein, Hao ~Wang, Thomas ~B{\"a}ck
``Neural Network Design: Learning from Neural Architecture Search"
\href{https://arxiv.org/abs/2011.00521}{[arXiv:2011.00521]}.

\bibitem{Roberts:2021fes}
D.~A.~Roberts, S.~Yaida and B.~Hanin,
``The Principles of Deep Learning Theory,''
\href{https://doi.org/10.1017/9781009023405}{Cambridge University Press, 2022,
ISBN 978-1-00-902340-5},
%doi:10.1017/9781009023405
\href{https://arxiv.org/abs/2106.10165}{[arXiv:2106.10165 [cs.LG]]}.

\bibitem{Alison:2019kud}
J.~Alison, S.~An, P.~Bryant, B.~Burkle, S.~Gleyzer, M.~Narain, M.~Paulini, B.~Poczos and E.~Usai,
``End-to-end particle and event identification at the Large Hadron Collider with CMS Open Data,''
\href{https://arxiv.org/abs/1910.07029}{[arXiv:1910.07029 [hep-ex]]}.


\bibitem{Chen:2022ddj}
C.~Chen, H.~Chen, W.~Q.~Niu and H.~Q.~Zheng,
``Identifying hadronic molecular states with a neural network,''
\href{https://doi.org/10.1140/epjc/s10052-023-11170-1}{Eur. Phys. J. C \textbf{83}, no.1, 52 (2023)},
%doi:10.1140/epjc/s10052-023-11170-1
\href{https://arxiv.org/abs/2205.03572}{[arXiv:2205.03572 [hep-ph]]}.

\bibitem{Andrews:2019faz}
M.~Andrews, J.~Alison, S.~An, P.~Bryant, B.~Burkle, S.~Gleyzer, M.~Narain, M.~Paulini, B.~Poczos and E.~Usai,
``End-to-end jet classification of quarks and gluons with the CMS Open Data,''
\href{https://doi.org/10.1016/j.nima.2020.164304}{Nucl. Instrum. Meth. A \textbf{977}, 164304 (2020)},
%doi:10.1016/j.nima.2020.164304
\href{https://arxiv.org/abs/1902.08276 }{[arXiv:1902.08276 [hep-ex]]}.

\bibitem{Zhang:2023czx}
Z.~Zhang, J.~Liu, J.~Hu, Q.~Wang and U.~G.~Mei\ss{}ner,
``Revealing the nature of hidden charm pentaquarks with machine learning,''
\href{https://doi.org/10.1016/j.scib.2023.04.018}{Sci. Bull. \textbf{68}, 981-989 (2023)},
%doi:10.1016/j.scib.2023.04.018
\href{https://arxiv.org/abs/2301.05364}{[arXiv:2301.05364 [hep-ph]]}.

\bibitem{Ng:2021ibr}
L.~Ng \textit{et al}. [Joint Physics Analysis Center and JPAC],
``Deep learning exotic hadrons,''
\href{https://doi.org/10.1103/PhysRevD.105.L091501}{Phys. Rev. D \textbf{105}, no.9, L091501 (2022)},
%doi:10.1103/PhysRevD.105.L091501
\href{https://arxiv.org/abs/2110.13742}{[arXiv:2110.13742 [hep-ph]]}.



\bibitem{ExaTrkX:2020nyf}
X.~Ju \textit{et al}. [Exa.TrkX],
``Graph Neural Networks for Particle Reconstruction in High Energy Physics detectors,''
\href{https://arxiv.org/abs/2003.11603}{[arXiv:2003.11603 [physics.ins-det]]}.


\bibitem{Bahtiyar:2022une}
H.~Bahtiyar,
``Predicting the masses of exotic hadrons with data augmentation using multilayer perceptron,''
\href{https://doi.org/10.1142/S0217751X23500033}{Int. J. Mod. Phys. A \textbf{38}, no.01, 2350003 (2023)},
%doi:10.1142/S0217751X23500033
\href{https://arxiv.org/abs/2208.09538}{[arXiv:2208.09538 [hep-ph]]}.

\bibitem{Gal:2020dyc}
Y.~Gal, V.~Jejjala, D.~K.~Mayorga Pe\~na and C.~Mishra,
``Baryons from Mesons: A Machine Learning Perspective,''
\href{https://doi.org/10.1142/S0217751X22500312}{Int. J. Mod. Phys. A \textbf{37}, no.06, 2250031 (2022)},
%doi:10.1142/S0217751X22500312
\href{https://arxiv.org/abs/2003.10445 }{[arXiv:2003.10445 [hep-ph]]}.


\bibitem{Zhang:2024bld}
Y.~Zhang, C.~Mo, X.~Chen, B.~Li, H.~Chen, J.~Hu and L.~Li,
``Search for Long-lived Particles at Future Lepton Colliders Using Deep Learning Techniques,''
\href{https://arxiv.org/abs/2401.05094}{[arXiv:2401.05094 [hep-ex]]}.


\bibitem{Bogatskiy:2023nnw}
A.~Bogatskiy, T.~Hoffman, D.~W.~Miller, J.~T.~Offermann and X.~Liu,
``Explainable Equivariant Neural Networks for Particle Physics: PELICAN,''
\href{https://arxiv.org/abs/2307.16506}{[arXiv:2307.16506 [hep-ph]]}.


\bibitem{Athanassopoulos:2003qe}
S.~Athanassopoulos, E.~Mavrommatis, K.~A.~Gernoth and J.~W.~Clark,
``Nuclear mass systematics using neural networks,''
\href{https://doi.org/10.1016/j.nuclphysa.2004.08.006}{Nucl. Phys. A \textbf{743}, 222-235 (2004)},
%doi:10.1016/j.nuclphysa.2004.08.006
\href{https://arxiv.org/abs/nucl-th/0307117}{[arXiv:nucl-th/0307117 [nucl-th]]}.




\bibitem{Arganda:2024eub}
E.~Arganda, M.~Epele, N.~I.~Mileo and R.~A.~Morales,
``Machine-Learning Performance on Higgs-Pair Production associated with Dark Matter at the LHC,''
\href{https://arxiv.org/abs/2401.03178}{[arXiv:2401.03178 [hep-ph]]}.


\bibitem{Chacon:2023nhw}
J.~Chac\'on, I.~G\'omez-Vargas, R.~M.~M\'endez and J.~A.~V\'azquez,
``Analysis of dark matter halo structure formation in N-body simulations with machine learning,''
\href{https://doi.org/10.1103/PhysRevD.107.123515}{Phys. Rev. D \textbf{107}, no.12, 123515 (2023)},
%doi:10.1103/PhysRevD.107.123515
\href{https://arxiv.org/abs/2303.09098}{[arXiv:2303.09098 [astro-ph.CO]]}.

\bibitem{Thais:2022iok}
S.~Thais, P.~Calafiura, G.~Chachamis, G.~DeZoort, J.~Duarte, S.~Ganguly, M.~Kagan, D.~Murnane, M.~S.~Neubauer and K.~Terao,
``Graph Neural Networks in Particle Physics: Implementations, Innovations, and Challenges,''
\href{https://arxiv.org/abs/2203.12852}{[arXiv:2203.12852 [hep-ex]]}.


\bibitem{Paganini:2017dwg}
M.~Paganini, L.~de Oliveira and B.~Nachman,
``CaloGAN : Simulating 3D high energy particle showers in multilayer electromagnetic calorimeters with generative adversarial networks,''
\href{https://doi.org/10.1103/PhysRevD.97.014021}{Phys. Rev. D \textbf{97}, no.1, 014021 (2018)},
%doi:10.1103/PhysRevD.97.014021
\href{https://arxiv.org/abs/1712.10321}{[arXiv:1712.10321 [hep-ex]]}.

\bibitem{Krause:2021ilc}
C.~Krause and D.~Shih,
``Fast and accurate simulations of calorimeter showers with normalizing flows,''
\href{https://doi.org/10.1103/PhysRevD.107.113003}{Phys. Rev. D \textbf{107}, no.11, 113003 (2023)},
%doi:10.1103/PhysRevD.107.113003
\href{https://arxiv.org/abs/2106.05285}{[arXiv:2106.05285 [physics.ins-det]]}.

\bibitem{Butter:2021csz}
A.~Butter, T.~Heimel, S.~Hummerich, T.~Krebs, T.~Plehn, A.~Rousselot and S.~Vent,
``Generative networks for precision enthusiasts,''
\href{https://doi.org/10.21468/SciPostPhys.14.4.078}{SciPost Phys. \textbf{14}, no.4, 078 (2023)},
%doi:10.21468/SciPostPhys.14.4.078
\href{https://arxiv.org/abs/2110.13632}{[arXiv:2110.13632 [hep-ph]]}.

\bibitem{Heimel:2022wyj}
T.~Heimel, R.~Winterhalder, A.~Butter, J.~Isaacson, C.~Krause, F.~Maltoni, O.~Mattelaer and T.~Plehn,
``MadNIS - Neural multi-channel importance sampling,''
\href{https://doi.org/10.21468/SciPostPhys.15.4.141}{SciPost Phys. \textbf{15}, no.4, 141 (2023)},
%doi:10.21468/SciPostPhys.15.4.141
\href{https://arxiv.org/abs/2212.06172 }{[arXiv:2212.06172 [hep-ph]]}.



\bibitem{Hashemi:2023ruu}
H.~Hashemi, N.~Hartmann, S.~Sharifzadeh, J.~Kahn and T.~Kuhr,
``Ultra-High-Resolution Detector Simulation with Intra-Event Aware GAN and Self-Supervised Relational Reasoning,''
\href{https://arxiv.org/abs/2303.08046}{[arXiv:2303.08046 [physics.ins-det]]}.


%\cite{Brambilla:2019esw}





\bibitem{Baldi:2018qhe}
P.~Baldi, J.~Bian, L.~Hertel and L.~Li,
``Improved Energy Reconstruction in NOvA with Regression Convolutional Neural Networks,''
\href{https://doi.org/10.1103/PhysRevD.99.012011}{Phys. Rev. D \textbf{99}, no.1, 012011 (2019)},
%doi:10.1103/PhysRevD.99.012011
\href{https://arxiv.org/abs/1811.04557}{[arXiv:1811.04557 [physics.ins-det]]}.

\bibitem{Karagiorgi:2022qnh}
G.~Karagiorgi, G.~Kasieczka, S.~Kravitz, B.~Nachman and D.~Shih,
``Machine learning in the search for new fundamental physics,''
\href{https://doi.org/10.1038/s42254-022-00455-1}{Nature Rev. Phys. \textbf{4}, no.6, 399-412 (2022)}.
%doi:10.1038/s42254-022-00455-1

\bibitem{Guerrieri:2022bxg}
G.~Guerrieri [ATLAS],
``Regression Deep Neural Networks for top-quark-pair resonance searches in the dilepton channel,''
\href{https://doi.org/10.1393/ncc/i2022-22110-0}{Nuovo Cim. C \textbf{45}, no.5, 110 (2022)}.
%doi:10.1393/ncc/i2022-22110-0



	
	
\bibitem{Keras}
Keras: Deep learning library for theano and tensorflow
F. Chollet

%\cite{Abadi:2016kxi}
\bibitem{Abadi:2016kxi}
M.~Abadi, P.~Barham, J.~Chen, Z.~Chen, A.~Davis, J.~Dean, M.~Devin, S.~Ghemawat, G.~Irving and M.~Isard, \textit{et al}.
``TensorFlow: A system for large-scale machine learning,''
\href{https://arxiv.org/abs/1605.08695}{[arXiv:1605.08695 [cs.DC]]}.
%81 citations counted in INSPIRE as of 20 Dec 2023









\bibitem{Brambilla:2019esw}
N.~Brambilla, S.~Eidelman, C.~Hanhart, A.~Nefediev, C.~P.~Shen, C.~E.~Thomas, A.~Vairo and C.~Z.~Yuan,
``The $XYZ$ states: experimental and theoretical status and perspectives,''
\href{https://doi.org/10.1016/j.physrep.2020.05.001}{Phys. Rept. \textbf{873} (2020), 1-154}
%doi:10.1016/j.physrep.2020.05.001
\href{https://arxiv.org/abs/1907.07583}{[arXiv:1907.07583 [hep-ex]].}
%658 citations counted in INSPIRE as of 25 Feb 2024

%\cite{Close:2002zu}
\bibitem{Close:2002zu}
F.~E.~Close and N.~A.~Tornqvist,
``Scalar mesons above and below 1-GeV,''
\href{https://doi.org/10.1088/0954-3899/28/10/201}{J. Phys. G \textbf{28} (2002), R249-R267},
\href{https://arxiv.org/abs/hep-ph/0204205 }{[arXiv:hep-ph/0204205 [hep-ph]].}
%446 citations counted in INSPIRE as of 25 Feb 2024


\bibitem{Aachen-Berlin-CERN-London-Vienna:1978zeh}
G.~Otter \textit{et al.} [Aachen-Berlin-CERN-London-Vienna],
``Partial Wave Analysis of the ($K \pi \pi$) System in the L Region in $K^- p \to (K^- \pi^+ \pi^-) p$ at 10-{GeV}/$c$, 14.3-{GeV}/$c$ and 16-{GeV}/$c$,''
\href{https://doi.org/10.1016/0550-3213(79)90411-5}{Nucl. Phys. B \textbf{147}, 1-14 (1979)}.
%doi:10.1016/0550-3213(79)90411-5



\bibitem{Taboada-Nieto:2022igy}
U.~Taboada-Nieto, P.~G.~Ortega, D.~R.~Entem, F.~Fern\'andez and J.~Segovia,
``Kaon spectrum revisited: bound states of high energy and spin,''
\href{https://doi.org/10.1140/epja/s10050-023-00963-3}{Eur. Phys. J. A \textbf{59}, no.3, 40 (2023)}.
%doi:10.1140/epja/s10050-023-00963-3
\href{https://arxiv.org/abs/2209.12555}{[arXiv:2209.12555 [hep-ph]]}.




%\cite{CrystalBarrel:2019zqh}
\bibitem{CrystalBarrel:2019zqh}
M.~Albrecht \textit{et al.} [Crystal Barrel],
``Coupled channel analysis of ${\bar{p}p}\,\rightarrow \,\pi ^0\pi ^0\eta $, ${\pi ^0\eta \eta }$ and ${K^+K^-\pi ^0}$ at 900 MeV/c and of ${\pi \pi }$-scattering data,''
\href{https://doi.org/10.1140/epjc/s10052-020-7930-x}{Eur. Phys. J. C \textbf{80} (2020) no.5, 453}
\href{https://arxiv.org/abs/1909.07091}{[arXiv:1909.07091 [hep-ex]].}
%25 citations counted in INSPIRE as of 25 Feb 2024

%\cite{Abachi:1988fw}
\bibitem{Abachi:1988fw}
S.~Abachi, C.~Akerlof, P.~Baringer, D.~Blockus, B.~Brabson, J.~M.~Brom, B.~G.~Bylsma, J.~Chapman, B.~Cork and R.~Debonte, \textit{et al.}
``Measurement of Upper Limits for the Decay Widths of $D^*$+ and $D^*_0$,''
\href{https://doi.org/10.1016/0370-2693(88)91811-4}{Phys. Lett. B \textbf{212} (1988), 533-536}.
%15 citations counted in INSPIRE as of 25 Feb 2024

%\cite{Asghar:2019qjl}
%\bibitem{Asghar:2019qjl}
%I.~Asghar, F.~Akram, B.~Masud and M.~A.~Sultan,
%``Properties of excited charmed-bottom mesons,''
%\href{https://doi.org/10.1103/PhysRevD.100.096002}{Phys. Rev. D \textbf{100}, no.9, 096002 (2019),}
%\href{https://arxiv.org/abs/1910.02680}{[arXiv:1910.02680 [hep-ph]].}
%11 citations counted in INSPIRE as of 07 Mar 2024

%\cite{Asghar:2023fvk}
%\bibitem{Asghar:2023fvk}
%I.~Asghar and N.~Akbar,
%``Spectrum and Decay Properties of Bottomonium Mesons,''
%\href{https://arxiv.org/abs/2309.15438}{[arXiv:2309.15438 [hep-ph]].}
%0 citations counted in INSPIRE as of 07 Mar 2024

%\cite{BaBar:2006eep}
\bibitem{BaBar:2006eep}
B.~Aubert \textit{et al.} [BaBar],
``A Study of the D*(sJ)(2317) and D(sJ)(2460) Mesons in Inclusive c anti-c Production Near (s)**(1/2) = 10.6-GeV,''
\href{https://doi.org/10.1103/PhysRevD.74.032007}{Phys. Rev. D \textbf{74} (2006), 032007}
\href{https://arxiv.org/abs/hep-ex/0604030}{[arXiv:hep-ex/0604030 [hep-ex]].}
%97 citations counted in INSPIRE as of 25 Feb 2024

%\cite{BESIII:2022yga}
\bibitem{BESIII:2022yga}
M.~Ablikim \textit{et al.} [BESIII],
``Observation of Resonance Structures in $e^+e^- \rightarrow \pi^+ \pi ^- \psi_2(3823)  $ and Mass Measurement of $ \psi_2(3823) $,''
\href{https://doi.org/10.1103/PhysRevLett.129.102003}{Phys. Rev. Lett. \textbf{129} (2022) no.10, 102003}
\href{https://arxiv.org/abs/2203.05815}{[arXiv:2203.05815 [hep-ex]].}
%17 citations counted in INSPIRE as of 25 Feb 2024

%\cite{Belle:2012fkf}
\bibitem{Belle:2012fkf}
R.~Mizuk \textit{et al.} [Belle],
``Evidence for the $\eta_b(2S)$ and observation of $h_b(1P) \to \eta_b(1S) \gamma$ and $h_b(2P) \to \eta_b(1S) \gamma$,''
\href{https://doi.org/10.1103/PhysRevLett.109.232002}{Phys. Rev. Lett. \textbf{109}, 232002 (2012),}
\href{https://arxiv.org/abs/1205.6351}{[arXiv:1205.6351 [hep-ex]].}
%126 citations counted in INSPIRE as of 07 Mar 2024

%\cite{Agaev:2018sco}
%\bibitem{Agaev:2018sco}
%S.~S.~Agaev, K.~Azizi and H.~Sundu,
%``The strong decays of the light scalar mesons $f_0(500)$ and $f_0(980)$,''
%\href{https://doi.org/10.1016/j.physletb.2018.07.042}{Phys. Lett. B \textbf{784}, 266-270 (2018),}
%\href{https://arxiv.org/abs/1804.01726}{[arXiv:1804.01726 [hep-ph]].}
%10 citations counted in INSPIRE as of 08 Mar 2024

%\cite{Agaev:2021jsz}
\bibitem{Agaev:2021jsz}
S.~S.~Agaev, K.~Azizi and H.~Sundu,
``Doubly charged vector tetraquark $Z_V^{++}=[cu][\bar{s}\bar{d}]$,''
\href{https://doi.org/10.1016/j.physletb.2021.136530}{Phys. Lett. B \textbf{820}, 136530 (2021),}
\href{https://arxiv.org/abs/2105.00081}{[arXiv:2105.00081 [hep-ph]].}
%14 citations counted in INSPIRE as of 08 Mar 2024





	
	
	
	


	

	

	
	
	
\end{thebibliography}
\end{document}